\documentclass{elsarticle}
\usepackage[utf8]{inputenc}
\usepackage[top=1in,bottom=1in,left=1in,right=1in]{geometry}
\usepackage{amssymb}
\usepackage{amsmath}
\usepackage{xcolor}
\usepackage{graphicx}
\usepackage{epstopdf}
\usepackage{subcaption}
\usepackage{braket}

\numberwithin{equation}{section}
\numberwithin{figure}{section}
\numberwithin{table}{section}
\newcommand{\vk}{{\bf k}}
\newcommand{\vq}{{\bf q}}
\newcommand{\vx}{{\bf x}}

\begin{document}
\begin{frontmatter}

\title{Analyzing and Predicting Non-equilibrium Many-body Dynamics via Dynamic Mode Decomposition}
\author[1]{Jia {Yin}}
\ead{jiayin@lbl.gov}
\author[2]{Yang-hao {Chan}}
\author[3]{Felipe da {Jornada}}
\author[4]{Diana {Qiu}}
\author[1]{Chao {Yang}\corref{cor1}}
\ead{cyang@lbl.gov}
\cortext[cor1]{Corresponding author: 
	Tel.: +1-510-486-6424;  
	fax: +1-510-486-5812;} 
\author[5,6]{Steven G. {Louie}}
\ead{sglouie@berkeley.edu}

\address[1]{Computational Research Division, Lawrence Berkeley National Laboratory, Berkeley, CA 94720, USA}
\address[2]{Institute of Atomic and Molecular Sciences, Academia Sinica, Taipei 10617, Taiwan}
\address[3]{Department of Materials Science and Engineering, Stanford University, Stanford, CA 94305, USA}
\address[4]{School of Engineering \& Applied Science, Yale University, New Haven, CT 06520, USA}
\address[5]{Department of Physics, University of California at Berkeley, Berkeley, CA 94720, USA}
\address[6]{Materials Sciences Division, Lawrence Berkeley National Laboratory, Berkeley, CA 94720, USA}

\begin{abstract}
    Simulating the dynamics of a nonequilibrium quantum many-body system by computing the two-time Green's function associated with such a system is computationally challenging.  However, we are often interested in the time diagonal of such a Green's function or time dependent physical observables that are functions of one time. In this paper, we discuss the possibility of using dynamic model decomposition (DMD), a data-driven model order reduction technique, to characterize one-time observables associated with the nonequilibrium dynamics using snapshots computed within a small time window. The DMD method allows us to efficiently predict long time dynamics from a limited number of trajectory samples. We demonstrate the effectiveness of DMD on a model two-band system. We show that, in the equilibrium limit, the DMD analysis yields results that are consistent with those produced from a linear response analysis. In the nonequilibrium case, the extrapolated dynamics produced by DMD is more accurate than a special Fourier extrapolation scheme presented in this paper. We point out a potential pitfall of the standard DMD method caused by insufficient spatial/momentum resolution of the discretization scheme. We show how this problem can be overcome by using a variant of the DMD method known as higher order DMD.\\
\end{abstract}

\begin{keyword}
	Dynamic mode decomposition, Koopman operator, Non-equilibrium quantum many-body dynamics, Kadanoff-Baym equations
\end{keyword}

\end{frontmatter}

\section{Introduction}\label{sec:intro}
%

Simulating a quantum many-body system away from equilibrium is a challenging
    task. Although time-dependent physical observables can be computed from the solution of a time-dependent Schr\"{o}dinger equation with a time-dependent Hamiltonian, such a brute-force approach is limited to small systems defined in a small dimensional Hilbert space. A more practical approach is to focus on the Green's function which is a two-point correlator of the creation and annhilation field operators defined on the Keldysh contour~\cite{kadanoff1962quantum,keldysh1965zhetf}. The equation of motion satisfied by the two-time Green's function is a set of nonlinear integro-differential equations~\cite{lipavsky1986generalized}. Evolving the Green's function numerically on a two-time grid is highly non-trivial, and the presence of the integral kernel in these equations makes both the memory requirement and computational cost high if the long-time behavior of a physical observable is to be examined.
 
In this paper, we show how the long-term characteristics of the physical observable can be analyzed and predicted using a model reduction technique -- the dynamic mode decomposition (DMD) ~\cite{kutz2016dynamic,DMD0,schmid2011applications,TuRowley}. The DMD method is a practical data-driven model reduction method first proposed by Schimid~\cite{DMD0} to analyze the dynamics of a nonlinear and high-dimensional system. It extracts the spatial modes associated with temporal oscillations with distinct frequencies and growth/decay rates from a few samples of the trajectory.  These spatial and temporal modes obtained from the DMD analysis of the dynamics within a limited time window can in turn be used to extrapolate and predict the dynamics on a much longer time scale.

One of the main advantages of DMD over other dimension reduction techniques such as 
the principal component analysis (PCA)~\cite{jolliffe2016principal,wold1987principal} and proper orthogonal decomposition (POD)~\cite{POD,semeraro2012analysis} is that DMD provides both the spatial and temporal modes at the same time.  Furthermore, the spatial modes obtained from the DMD analysis is often more physical than the eigenvectors or singular vectors produced from PCA and POD.

To use DMD to predict long time behavior of certain physical observables (such as density) associated with the evolution of a many-body system out of equilibrium, we first solve the equation of motion satisfied by the two-time Green's function, i.e. the Kadanoff-Baym equation within a small time window, and perform DMD analysis on the one-time physical observables that can be obtained from the Green's function. 

Our paper is organized as follows. In section~\ref{sec:kbe}, we describe the model problem we use to demonstrate the effectiveness of the DMD method and the equation of motion satisfied by the two-time Green's function as well as one-time physical observable in both the equilibrium and non-equilibrium regimes.
The mathematical foundation of the DMD analysis and the numerical procedure for performing such an analysis is presented in section~\ref{sec:dmd}. In section~\ref{sec:hodmd}, we point out a potential problem of the DMD caused by an insufficient resolution in the spatial (or momentum) discretization of the state variable. We explain how this problem can be fixed by using a high-order DMD (HODMD) analysis which can be interpreted as time-delayed embedding of a nonlinear dynamical system.

The effectiveness of the DMD and HODMD procedures are reported and discussed in section~\ref{sec:examples}. In particular, we demonstrate that, in the equilibrium limit when linear response analysis can be performed, the DMD modes obtained from a real-time, time-dependent Hartree-Fock (TD-HF) simulation match well the eigenvectors obtained from solving the Bethe-Salpeter in the Kohn-Sham basis. This agreement also appears to hold for weakly non-equilibrium dynamics driven by a low intensity field. 

As the intensity of the driving field increases, the linear response theory does not hold.
To validate DMD and HODMD results, we compare the DMD and HODMD modes with spatial and temporal modes identified by performing a Fourier analysis of the observable trajectory and show their differences. 
To demonstrate that HODMD modes are more relevant and meaningful, we compare the extrapolated trajectories produced by HODMD and a modified Fourier scheme that tries to recover the decay rate by solving a nonlinear optimization problem. Our numerical results show that the HODMD extrapolation is much more accurate than the Fourier extrapolation, and the HODMD procedure is numerically more stable than the modified Fourier extrapolation scheme.


\section{The model problem and the Keldyish formalism}
\label{sec:kbe}

In this work, we focus on the dynamics of a simple two-band system, which exemplifies the semiconductor driven by  an external light field~\cite{perfetto2019pump}. The Hamiltonian consists of a time independent component $H_s$ that describes the many-body interaction as well as an external 
time dependent component that describes the light-matter coupling.

The system Hamiltonian has the form
\begin{align}
    H_s=\sum_\vk(\epsilon_{v\vk}c^\dagger_{v\vk} c_{v\vk} + \epsilon_{c\vk}c^\dagger_{c\vk}c_{c\vk})-U\sum_k c^\dagger_{c\vk}c_{c\vk}+\frac{U}{N}\sum_{\vk_1,\vk_2,\vq}c^\dagger_{v\vk_1+\vq}c^\dagger_{c\vk_2-\vq}c_{c\vk_2}c_{v\vk_1},
    \label{eq:ham}
\end{align}
where $\epsilon_{v\vk}$ ($\epsilon_{c\vk}$) is the band energy of the valence (conduction) band with momentum $\vk$, $U$ is the on-site interaction between the two bands, and $N$ is the number of sites in the system. The energy dispersion is given by \begin{align*}
    \epsilon_{v\vk}&=-2(1-\cos(\vk)) - E_g/2 \\
    \epsilon_{c\vk}&=2(1-\cos(\vk)) + E_g/2,
\end{align*}
with $E_g=1$ as the band gap. 

The light-matter coupling within the dipole approximation is
\begin{align}
    H_{ext}(t) = E(t)\sum_{\vk}(d_\vk c^\dagger_{c\vk}c_{v\vk}+d^*_\vk c^\dagger_{v\vk}c_{c\vk}),
    \label{eq:dipole}
\end{align}
where $E(t)$ is a time-dependent intensity of the field, and $d_\vk$ is the dipole matrix element. For simplicity we set $d_\vk=1$.
\eqref{eq:ham} together with \eqref{eq:dipole} describes how electrons and holes interact with each other and with a classical light field.

Although all time-dependent physical obserables can be obtained from the solution to the time-dependent Schr\"{o}dinger's equation
\begin{equation}
    i \hbar \frac{d\ket{\Psi(t)}}{dt} = H(t)\ket{\Psi(t)}, \ \ \rm{with} \ \ \ket{\Psi(0)} = \ket{\Psi_0},
    \label{eq:schrod}
\end{equation}
where $H(t) = H_s + H_{ext}(t)$ and $\Psi_0$ 
is the initial state of the wavefunction $\Psi$ at $t=0$,
the many-body nature of \eqref{eq:ham} renders the full solution of \eqref{eq:schrod} difficult. The computational complexity of the exact numerical solution grows exponentially with the system size. 

Since in most cases we are interested only in single particle physical observables, we apply the nonequilibrium Green's function (NEGF) approach~\cite{kadanoff1962quantum} to map the dynamics of the many-body system to the two-time Greens function $G_{i,j}(t,t') = -\frac{i}{\hbar} \langle T_{\mathcal{C}} \hat{c}_i(t) \hat{c}_j^{\dagger}(t')\rangle$,  where $\hat{c}_i(t)$ and $\hat{c}_j^\dagger(t')$ are annihilation and creation operators in the Heisenberg picture, $\mathcal{T}$ is the time-ordering operator, and the expectation is evaluated along the Keldysh contour~\cite{keldysh1965zhetf}. Since the model problem we focus on in this work consists of two bands, and the Green's function of interest is $G_{c,v}(t,t')$, we will drop the band indices below and simply denote the Green's function by $G(t,t')$.

It follows from the many-body perturbation theory that $G(t,t')$ satisfies 
the following equation of motion
\begin{align}
    \left[i\frac{d}{dt}-H(t)\right]G(t,t') = \delta(t,t') + \int_C \Sigma(t,\Bar{t})G(\Bar{t},t')d\Bar{t},
    \label{eq:KBE}
\end{align}
where $H(t)$ is now a single-particle Hamiltonian that includes the contribution of a time-dependent driving field, and
$\Sigma(t,t')$ is the self-energy that describes the many-body interaction. Equation~\eqref{eq:KBE} is accompanied by an ajoint equation which describes the time evolution over $t'$. These equations are coupled nonlinear integral differential equations that are collectively called the Kadanoff-Baym equations (KBE)~\cite{kadanoff1962quantum}. 
They must be solved numerically. Once \eqref{eq:KBE} and its ajoint equation are solved, the single particle physical observables can be computed through the relation between the density matrix and the time-diagonal part of the lesser Green's function, $\rho(t) = -iG^<(t,t)$.

In the NEGF approach, many-body interaction is captured by the self-energy term. Depending on the physical problem, a proper choice of self-energy is essential. In this work, we will use the Hartree-Fock (HF) and the second Born (2B) approximation of the self-energy, which can capture exciton physics and in addition, carrier scatterings respectively~\cite{Stefanucci2013}.

Due to the presence of the integral term in \eqref{eq:KBE} and its adjoint equation, the numerical solution of these coupled integral differential equations is nontrivial. 
Depending on the choice of the self-energy, the right-hand side of \eqref{eq:KBE} may be a nonlinear function of the two-time Green's function $G(t,t')$. As a result, each time evolution step would require solving a nonlinear system of equations. The computational complexity of solving the two-time KBE scales as $O(t^3)$ in the worst case. This high complexity severely limits its application beyond HF approximation. 

However, because the physical 
observable we are interested in, e.g., $\rho(t)$, is often a function of $t$ only, it may be possible to use a data-driven model order reduction technique to characterize the spatial and long-time temporal features of the dynamics satisfied by the one-time physical observe from samples of the observables sampled from a small time window. These samples are computed from the the numerical solution of the two-time KBE.  

In principle, a one-time physical observable satisfies a one-time equation of motion. For example, $\rho(t)$ is the solution to a differential equation of the form
\begin{equation}
   \frac{d}{dt}\rho(t) = f[\rho(t),t], 
   \label{eq:rhoeq}
\end{equation}
where $f[\rho(t),t]$ may be a complicated nonlinear function of $\rho(t)$ and $t$ for which an explicit analytic form is unknown.



For the model problem we will focus on in this work, we can write down the equation of motion for $\rho$ explicitly. 
We start from the equation of motion
\begin{align}
    i\frac{d}{d t} \rho(t) = \left [ H(t), \rho(t)\right],
    \label{eq:singlerhoeq}
\end{align}
where $H(t)=H_s+H_{ext}(t)$ is the total Hamiltonian.
The matrix element of density matrix is defined as the expectation value $\rho_{cv, \mathbf{k}}=\langle c^\dagger_{v\mathbf{k}}c_{c\mathbf{k}}\rangle$. In general Eq.~\ref{eq:singlerhoeq} couples to the density matrix with higher particle numbers and is not closed.
We can close the equation of motion by taking the HF approximation of the interaction term and obtain
\begin{align}
    i\frac{d}{d t} \rho_{cv,\mathbf{k}}(t) = (\epsilon_{v\mathbf{k}}-\epsilon_{c\mathbf{k}})\rho_{cv,\mathbf{k}}(t) + (f_{c\mathbf{k}}-f_{v\mathbf{k}})\left[E(t) - \frac{U}{N}\sum_\mathbf{k'}\rho_{cv,\mathbf{k'}}\right],
\end{align}
where $f_{c\mathbf{k}}$ and $f_{v\mathbf{k}}$ are the occupation numbers of conduction and valence bands respectively. In the weak field limit, $f_{c\mathbf{k}}=0$ and $f_{v\mathbf{k}}=1$. As a result, the off-diagonal matrix  element of $\rho(t)$ decouples from the diagonal terms to yield
\begin{align}
    i\frac{d}{d t} \rho_{cv,\mathbf{k}}(t)  +(\epsilon_{c\mathbf{k}}-\epsilon_{v\mathbf{k}})\rho_{cv,\mathbf{k}}
    -\frac{U}{N}\sum_\mathbf{k'}\rho_{cv,\mathbf{k'}} 
    = -E(t). 
    \label{eq:tdhfdriven}
\end{align}
When $E(t) = 0$ or when $|E(t)|$ is small, the solution of \eqref{eq:tdhfdriven} can be expressed in terms of the eigenvalues and eigenvectors of the Hamiltonian
\begin{align}
    H_{cv\mathbf{k},cv\mathbf{k'}} = (\epsilon_{c\mathbf{k}}-\epsilon_{v\mathbf{k}})\delta_{\mathbf{k}\mathbf{k'}} - \frac{U}{N}.
    \label{eq:BSEham}
\end{align}
This is the Bethe-Salpeter linear response Hamiltonian~\cite{attaccalite2011real}. We will use the eigenvalues and eigenvectors of BSE Hamiltonian \eqref{eq:BSEham} to validate the spatial and temporal features of $\rho_{cv,\mathbf{k}}$ obtained from a data driven reduced order model to be presented below.
\section{Dynamic mode decomposition}
\label{sec:dmd}

In this section, we briefly describe the basic principles of dynamic mode decomposition and the numerical procedure we use to perform this decomposition.

Dynamic mode decomposition (DMD) is a data-driven dimension reduction 
technique that can be used to extract important spatial and temporal features of a nonlinear dynamical system with a large number of degree of freedoms~\cite{kutz2016dynamic,mohan2018data,DMD0,towne2018spectral}. Future states of the nonlinear system can be predicted based on the extracted modes and frequencies.

Consider a dynamical system described by a nonlinear ordinary differential equation of the form 
\begin{equation}\label{eq:model}
    \frac{d\mathbf{x}(t)}{dt} = \mathbf{f}(\mathbf{x}(t), t), \quad t\geq 0,
\end{equation}
where $\mathbf{x}(t)\in\mathbb{C}^n$ is a time-dependent state variable, and $\mathbf{f}: \mathbb{C}^n\otimes \mathbb{R}^+ \rightarrow \mathbb{C}^n$ is a nonlinear function of $\mathbf{x}$ and time $t$. The goal of DMD is to identify a set of time independent spatial modes $\phi_1$, $\phi_2$, ... $\phi_k$ and a set of 
frequencies $\omega_1$, $\omega_2$, ... $\omega_k$ so that $\mathbf{x}(t)$ 
can be well-approximated by 
\begin{equation}
\mathbf{x}(t) \approx \sum_{\ell=1}^r \beta_\ell \phi_\ell e^{i\omega_\ell t},
\label{eq:xtexpand}
\end{equation}
where $\beta_\ell$'s are set of coefficients, and $r$ is relatively small.

The general strategy for obtaining the dynamic modes $\phi_\ell$ and corresponding frequencies $\omega_\ell$ is 
to map the trajectory of the nonlinear dynamics to the state of an infinite-dimensional
linear system that can easily be characterized via a spectral decomposition of the
linear operator that defines such a linear system. This strategy follows from the Koopman theory~\cite{Koopman1, Koopman2,takeishi2017learning} for reduce order modeling \cite{Bagheri1, Bagheri2, Mezic, Rowley1}.

In practice, we do not have the trajectory $\mathbf{x}(t)$ before \eqref{eq:model}
is solved. Yet, our hope is that the most important $\phi_i$'s and $\omega_i$'s can be obtained
by analyzing a small set of snapshots (or samples) of $\mathbf{x}(t)$ that 
we can solve.

Suppose snapshots of $\mathbf{x}(t)$ are available at $t_j = t_1+(j-1)\Delta t$, where $j=1,...,m$ and $\Delta t$ is a small time interval. We denote these snapshots by $\mathbf{x}_j = \mathbf{x}(t_j)$, and define
\[
\mathbf{X} = \left( \mathbf{x}_1 \: \mathbf{x}_2 \: \cdots \: \mathbf{x}_m \right).
\]

It follows from Koopman's theory that, in the limit of $\Delta t \rightarrow 0$ and $m \rightarrow \infty$, there exists an infinite-dimensional operator $\mathcal{A}$ 
such that 
\begin{equation}
    \mathcal{A} \mathbf{X} = \mathbf{X} \mathbf{S},
\label{eq:Koopman}
\end{equation}
where 
\begin{equation}
\mathbf{S} = 
\left(
\begin{array}{cccc}
  0  & \hdots & 0 & c_1  \\
  1  &        & 0 & c_2  \\
  \vdots & \vdots & \vdots & \vdots \\
  0  & \hdots & 1 & c_{m}
\end{array}
\right),
\end{equation}
where $c_j$ ($j=1,2,...,m$) is a set of coefficients \cite{DMDtoKoop}.

When $m$ is finite, \eqref{eq:Koopman} may not hold. In particular, we may not find a closure defined by the last column of $\mathbf{S}$. However, it is possible to construct a finite dimensional approximation to 
$\mathcal{A}$, denoted by $\mathbf{A}$ that minimizes
the difference between the leading $m$ columns between the left and right hand sides of \eqref{eq:Koopman}. 
To simplify notation, let us define
\begin{equation}
\mathbf{R} = \mathbf{A}\mathbf{X}_1 -\mathbf{X}_2
\label{eq:resid}
\end{equation}
where 
\begin{equation}
\mathbf{X}_1 =\left( \mathbf{x}_1 \: \mathbf{x}_2 \: \cdots \: \mathbf{x}_{m-1} \right) \ \ \mbox{and} \ \
\mathbf{X}_2 =\left( \mathbf{x}_2 \: \mathbf{x}_3 \: \cdots \: \mathbf{x}_{m} \right).
\label{eq:mats}
\end{equation}

It is easy to show that the minimizer of $\|\mathbf{R}\|_F$ is 
\begin{equation}\label{eq:A}
\mathbf{A} = \mathbf{X}_2  \mathbf{X}_1^{\dagger},
\end{equation}
where $\mathbf{X}_1^\dagger$ denotes the Moore-Penrose pseudoinverse of $\mathbf{X}_1$, which can be obtained from the singular value decomposition (SVD) \cite{SVD} of $\mathbf{X}_1$, i.e.
\begin{equation}\label{eq:SVD}
\mathbf{X}_1 = \mathbf{U\Sigma V}^*,
\end{equation}
where $\mathbf{U}\in\mathbb{C}^{n\times n}$, $\mathbf{\Sigma}\in\mathbb{C}^{n\times m}$, and $\mathbf{V}\in\mathbb{C}^{m\times m}$, $\mathbf{U}^{*}\mathbf{U} = \mathbf{I}$ and $\mathbf{V}^* \mathbf{V} = \mathbf{I}$.

For many large-scale problems, the snapshots contained in $\mathbf{X}_1$ may have a low rank $r \ll \min\{n,m\}$, 
i.e., the singular values on the 
diagonal of $\mathbf{\Sigma}$ decay rapidly. In this case, important dynamic modes can be obtained by projecting $\mathbf{A}$ into the subspace spanned by the leading right singular vectors of $\mathbf{A}$.

Take
\begin{equation}\label{eq:low_rank}
\widetilde{\mathbf{U}} = \mathbf{U}(:, 1:r), \quad \widetilde{\mathbf{\Sigma}} = \mathbf{\Sigma}(1:r, 1:r), \quad \widetilde{\mathbf{V}} = \mathbf{V}(:, 1:r),
\end{equation}
then $\widetilde{\mathbf{U}}\widetilde{\mathbf{\Sigma}}\widetilde{\mathbf{V}}^*$ projects $\mathbf{X}_1$ onto an $r$-dimensional subspace. Substituting
$X_1 \approx\widetilde{\mathbf{U}}\widetilde{\mathbf{\Sigma}}\widetilde{\mathbf{V}}^*$ into \eqref{eq:A} yields a rank-$r$ estimation of $\mathbf{A}$, i.e., 
\begin{equation}
\widetilde{\mathbf{A}} = \widetilde{\mathbf{U}}^*\mathbf{X}_2\widetilde{\mathbf{V}}\widetilde{\mathbf{\Sigma}}^{-1}\widetilde{\mathbf{U}}^*\widetilde{\mathbf{U}} = \widetilde{\mathbf{U}}^*\mathbf{X}_2\widetilde{\mathbf{V}}\widetilde{\mathbf{\Sigma}}^{-1}.
\label{eq:projA}
\end{equation}
To propagate the original system \eqref{eq:model}, what remains to be solved is the eigenvalue problem
\begin{equation}
\widetilde{\mathbf{A}}\mathbf{W} = \mathbf{W}\mathbf{\Lambda},
\label{eq:dmdev}
\end{equation}
where 
\begin{equation}
\mathbf{\Lambda} = \begin{bmatrix}
\lambda_1 & & \\
 & \ddots & \\
 & & \lambda_r
\end{bmatrix}
\end{equation}
is composed of the eigenvalues, and the columns of $\mathbf{W}$ give the corresponding eigenvectors. To obtain spectral modes in the original state space of $\mathbb{C}^n$, we perform the transformation 
\begin{equation}
\mathbf{\Phi} = \mathbf{X}_2\widetilde{\mathbf{V}}\widetilde{\mathbf{\Sigma}}^{-1}\mathbf{W}.
\label{eq:dmdmodes}
\end{equation}
The columns of $\mathbf{\Phi}$ are called the DMD modes. Denote
\begin{equation}
\mathbf{\Omega} = \frac{\ln{\mathbf{\Lambda}}}{\Delta t} = 
\begin{bmatrix}
i\omega_1^{\text{DMD}} & & \\
& \ddots & \\
& & i\omega_r^{\text{DMD}}
\end{bmatrix}, \quad \omega_\ell^{\text{DMD}} = -i\frac{\ln{\lambda_\ell}}{\Delta t}, \quad \ell = 1, ..., r,
\label{eq:dmdfreq}
\end{equation}
then the dynamics of $\mathbf{x}$ can be expressed as
\begin{equation}\label{eq:evol_DMD}
\mathbf{x}(t)\approx \sum_{\ell=1}^r\mathbf{\phi}_\ell\exp(i\omega_\ell^{\text{DMD}} t)b_\ell = \mathbf{\Phi}\exp(\Omega t)\mathbf{b}.
\end{equation}
In the expression, the amplitude vector $\mathbf{b}:=[b_1, ..., b_r]^T$ is taken to be the projection of initial value on to the DMD modes as
\begin{equation}\label{eq:b1}
\mathbf{b} = \mathbf{\Phi}^\dagger \mathbf{x}_1,
\end{equation}
or the least squares fit of \eqref{eq:evol_DMD} on the sampled trajectories:
\begin{equation}\label{eq:b2}
\mathbf{b} = \arg\min_{\tilde{\mathbf{b}}\in\mathbb{C}^n}\sum_{j=1}^m\|\mathbf{\Phi}\exp(\Omega t_j)\tilde{\mathbf{b}}-\vx_j\|_{l^2}^2,
\end{equation}
where $\|\cdot\|_{l^2}$ denotes the standard Euclidean norm of a vector. This completes the procedures of DMD.

From the flow of DMD, it is straightforward to see that the major computational cost comes from SVD \eqref{eq:SVD}, which is $O(\min(m^2n, mn^2))$.

As mentioned in Section \ref{sec:intro}, DMD is an equation-free, data-driven method. There is no need to know about the underlying dynamics function $\mathbf{f}(\mathbf{x}(t), t)$ in \eqref{eq:model}. Based on the data from first few time steps, it is possible to predict future states of the system. Moreover, it only focuses on the principal $r$ dimensions instead of the overall $n$ dimensions of the state in order to reduce computational cost. As a result, the method is probably of great value for many complicated nonlinear, or high-dimensional dynamical systems.

\section{Higher order dynamic mode decomposition}\label{sec:hodmd}
The number of spatial (momentum) and temporal modes $r$ in \eqref{eq:evol_DMD} is determined by the dimension of the projected Koopman operator $\tilde{\mathbf{A}}$ defined in ~\eqref{eq:projA}, which is projected from the approximate Koopman operator \eqref{eq:A} that maps $\mathbf{X}_1$ to $\mathbf{X}_2$.
When the snapshots $\mathbf{x}_j$ are discretized on 
a small number of spatial or momentum grid points, the dimension of $\mathbf{A}$, and consequently the dimension of $\tilde{\mathbf{A}}$ may be too small to accommodate the number of spatial and temporal modes present in the true dynamics of $\mathbf{x}(t)$.

This problem can possibly be resolved by using a finer spatial or momentum discretization scheme to increase the dimension of $\mathbf{A}$ and $\tilde{\mathbf{A}}$. However, this would inevitably increase the cost for generating the time snapshots for performing the DMD analysis. It is not clear, a priori, how fine a spatial or momentum grid one needs to resolve all significant spatial and temporal modes in the true dynamics satisfied by $\mathbf{x}(t)$.

Fortunately, this problem can be addressed by using the technique of time-delay embedding~\cite{broomhead1986extracting,packard1980geometry,Pan2020,Taken} to construct a better approximation to the Koopman operator without increasing number of spatial or momentum grid points in $\mathbf{x}_j$.

The key observation used in time-delay embedding can be described as follows. Let us partition a snapshot $\mathbf{x}$ discretized on a fine spatial or momentum grid as $\mathbf{x} = 
(\mathbf{x}_c, \mathbf{x}_f)^T$, where 
$\mathbf{x}_c$ corresponds to a subset of $\mathbf{x}$ defined on a (coarse) subset of grid points. If $\mathbf{A}$ is the Koopman operator that maps $\mathbf{x}(t_d)$ to
$\mathbf{x}(t_{d+1})$, i.e.
\begin{equation}\label{eq:Koopman2}
    \left[
    \begin{array}{c}
    \mathbf{x}_c(t_{d+1}) \\
    \mathbf{x}_f(t_{d+1}) \\
    \end{array}
    \right]
    =
    \left[
    \begin{array}{cc}
    \mathbf{A}_{11} & \mathbf{A}_{12}\\
    \mathbf{A}_{21} & \mathbf{A}_{22}\\
    \end{array}
    \right]
    \left[
    \begin{array}{c}
    \mathbf{x}_c(t_{d}) \\
    \mathbf{x}_f(t_{d}) \\
    \end{array}
    \right],
\end{equation}
where $\mathbf{A}$ is partitioned conformally with the partition of $\mathbf{x}$, then it is easy
to show that
\begin{equation}
    \mathbf{x}_c(t_{d+1})
    = \mathbf{A}_{11}\mathbf{x}_c(t_d)
    + \sum_{j=0}^{d-2} \mathbf{A}_{12}\mathbf{A}_{22}^j\mathbf{A}_{21} \mathbf{x}_c(t_{d-j-1}) 
    + \mathbf{A}_{12}\mathbf{A}_{22}^{d-1} \mathbf{x}_f(t_1).
    \label{eq:Aunfold}
\end{equation}

If the last term in \eqref{eq:Aunfold} is negligibly small, we can represent a coarsely sampled state $\mathbf{x}_c(t_{d+1})$ as a linear combination of time delayed states 
$\mathbf{x}_c(t_{d-j})$ for $j = 0,1,2,...,d-1$.  If this relationship holds 
for all $t$, we can construct an augmented Koopman operator 
\begin{equation}
    \tilde{\mathbf{C}} = \begin{bmatrix}
        \mathbf{0} & \mathbf{I} & \mathbf{0} & ... & \mathbf{0} & \mathbf{0}\\
        \mathbf{0} & \mathbf{0} & \mathbf{I} & ... & \mathbf{0} & \mathbf{0}\\
        ... & ... & ... & ... & ... & ...\\
        \mathbf{0} & \mathbf{0} & \mathbf{0} & ... & \mathbf{I} & \mathbf{0}\\
        \tilde{\bf B}_1 & \tilde{\bf B}_2 & \tilde{\bf B}_3 & ... & \tilde{\bf B}_{d-1} & \tilde{\bf B}_d
    \end{bmatrix},
\end{equation}
that maps $\tilde{\vx}_j$ to $\tilde{\vx}_{j+1}$ where $\tilde{\mathbf{B}}_1 = \mathbf{A}_{12}\mathbf{A}_{22}^{d-2}(\mathbf{A}_{21}+\mathbf{A}_{22})$, $\tilde{\mathbf{B}}_j = \mathbf{A}_{12}\mathbf{A}_{22}^{d-j-1}\mathbf{A}_{21}$, $j=2,...,d-1$, $\tilde{\mathbf{B}}_d = \mathbf{A}_{11}$, and  
\begin{equation}\label{eq:aug_x}
    \tilde{\vx}_j = \begin{bmatrix}
        \vx_j\\ \vx_{j+1}\\ ...\\ \vx_{j+d-2}\\ \vx_{j+d-1}
    \end{bmatrix}.
\end{equation}

An approximation to $\tilde{\mathbf{C}}$ can then be obtained by solving the least squares problem
\begin{equation}
\min_{\tilde{\mathbf{C}}} \| \tilde{\mathbf{C}} \tilde{\mathbf{X}}_1 
- \tilde{\mathbf{X}}_2
\|_F,
\label{eq:minC}
\end{equation}
where the snapshot matrices $\tilde{\mathbf{X}}_1$ and $\tilde{\mathbf{X}}_2$ are defined as 
\[
\tilde{\mathbf{X}}_1 = 
\left(\tilde{\vx}_1 \:
\tilde{\vx}_2 \: \cdots \:  \tilde{\vx}_{m-d} \right), \ \ \mbox{and} \ \ \tilde{\mathbf{X}}_2 = 
\left(\tilde{\vx}_2 \:
\tilde{\vx}_3 \: \cdots \:  \tilde{\vx}_{m-d+1} \right),
\]
where $m$ is the total number of sampled snapshots of $\mathbf{x}(t)$.

When $\tilde{\mathbf{X}}_1$ is low rank, the solution to \eqref{eq:minC} can be approximated from a subspace defined by the singular vectors associated with dominant singular vectors of $\tilde{\mathbf{X}}_1$ using the same procedure described in the previous section.
This modified procedure yields the higher order dynamic mode decomposition (HODMD) described in \cite{HODMD}. 

When each column of $\mathbf{\tilde{X}}_1$ consists of the concatenation of $d$ consecutive snapshots,  each spatial HODMD mode is a vector of length $nd$. To reconstruct or extrapolate the trajectory of $\mathbf{x}(t)$ by \eqref{eq:evol_DMD}, we take $\phi_{\ell}$ to be the first $n$ elements of the $\ell$th spatial HODMD mode.

Because the number of rows in the snapshot matrices $\mathbf{\tilde{X}}_1$ and $\mathbf{\tilde{X}}_2$ used in HODMD can be much larger than those in a standard DMD, the computational cost of HODMD is generally higher. Furthermore, when $\Delta t$ is relatively small, columns of the snapshot matrix can become more linearly dependent.  Although this problem can in principle be resolved by the truncated SVD performed in \eqref{eq:low_rank}, sometimes it may be difficult to choose an optimal singular value cutoff threshold for truncation.  To reduce the computational cost and the level of linear dependency among columns of $\mathbf{\tilde{X}_1}$, we can increase the temporal distance between the augmented snapshots in $\mathbf{\tilde{X}}_1$ and $\mathbf{\tilde{X}}_2$. For example, we can define them as
\begin{equation}\label{eq:DMDc-d}
	\mathbf{\tilde{X}}_1 = \begin{bmatrix}
		\mathbf{x}_1 & \mathbf{x}_{s+1} & ... & \mathbf{x}_{ps+1}\\
		\mathbf{x}_2 & \mathbf{x}_{s+2} & ... & \mathbf{x}_{ps+2}\\
		\vdots & \vdots & \vdots & \vdots\\
		\mathbf{x}_{d} & \mathbf{x}_{s+d} & ... & \mathbf{x}_{ps+d}
	\end{bmatrix}, \quad
	\mathbf{\tilde{X}}_2 = \begin{bmatrix}
		\mathbf{x}_{2} & \mathbf{x}_{s+2} & ... & \mathbf{x}_{ps+2}\\
		\mathbf{x}_{3} & \mathbf{x}_{s+3} & ... & \mathbf{x}_{ps+3}\\
		\vdots & \vdots & \vdots & \vdots\\
		\mathbf{x}_{d+1} & \mathbf{x}_{s+d+1} & ... & \mathbf{x}_{ps+d+1}
	\end{bmatrix},
\end{equation}
where $p$ and $s$ are some integers that satisfy $ps+d+1 \leq m$. A HODMD method associated with the parameters $d$ and $s$ will be denoted by HODMD($d$,$s$).

%

\section{Results and discussions}\label{sec:examples}

In this section, we give some examples on how
to use DMD to extract spatial and temporal modes of the dynamics associated with the simple 
two-band system defined by \eqref{eq:ham} when it is driven by a time-dependent field $E(t)=I\delta(t)$ through the light-matter interaction term \eqref{eq:dipole}. As we indicated in section~\ref{sec:kbe}, instead of solving the many-body problem directly, we use a NEGF formalism to compute a single particle Green's function by solving the KBE \eqref{eq:KBE} and its adjoint equation. As this is a two-band system, if we take $n$ $k$-points, and evolve the system for $m$ time steps, then the results of Green's function $G$ forms an $n\times 2\times 2\times m\times m$ matrix. From the solutions, the density matrix $\rho$ is obtained by $\rho(\cdot,\cdot,\cdot, t) = -iG^<(\cdot,\cdot,\cdot, t, t)$, which gives an $n\times 2\times 2\times m$ matrix. We use DMD to analyze and 
predict the dynamics of $\rho$ with the second and third indices fixed by 1 and 2, respectively. The data of $\rho(k, t)$ can thus be seen as an $n$-by-$m$ matrix, and we denote the entries by
\begin{equation}
\rho_{s,j} = \rho(k_s, t_j), \quad s=1, ..., n, \; j=1, ..., m
\end{equation}
for simplicity, where
\begin{equation}
    k_s = -\pi+2(s-1)\pi/n, \quad t_j = t_1+(j-1)\Delta t, \quad s=1, ..., n, \; j=1, ..., m.
\end{equation}
Each snapshot can be represented as
\begin{equation}
    \vx_j = [\rho_{1, j}, \rho_{2, j}, ..., \rho_{n, j}]^T, \quad j=1, ..., m,
\end{equation}
where $T$ stands for the transpose of a matrix. The data matrices $\mathbf{X}_1$ and $\mathbf{X}_2$ are then constructed through \eqref{eq:mats}.

We consider both the HF and second Born self-energies in \eqref{eq:KBE}. We also test DMD for different levels of field intensity $I$.
In all cases, we compare the DMD modes with spectral modes obtained from the Fourier analysis of the density trajectory.

In the weak intensity limit, it is known that we can perform a linear response analysis to obtain the spectral modes of the dynamics by solving the Casida equation or the Bethe-Salpeter equations (BSE) for two-particle neutral excitations.  In this regime, we can compare the modes extracted by DMD with the BSE eigenvectors. On the other hand, when $E(t)$ is sufficiently large, linear reponse can no longer accurately capture the dynamics of the Green's function whereas DMD can still be performed because it is designed to analyze nonlinear dynamics. 

\subsection{KBE with Hatree-Fock self-energy approximation}
When $E(t)=0$ and the self-energy term $\Sigma(t,t')$ is chosen to be the HF approximation, which is static,
the HF self-energy term can be absorbed into the single particle
Hamiltonian. As a result, the KBE reduces to time-dependent Hartree-Fock equations.
We solved this time-dependent problem by using a second-order Runge-Kutta integrator within the time interval $[0,500]$,  with a time step of $\Delta t = 0.1$.  Note that the time unit is defined as 1/energy unit. We have not assign specific unit to either the time or energy. Four $k$-points are sampled in the Brillouin zone, i.e., $n=4$. Therefore, each snapshot of the data matrices is a vector with 4 elements. 

\begin{figure}[ht]
    	\centering
    	\includegraphics[width=0.45\textwidth]{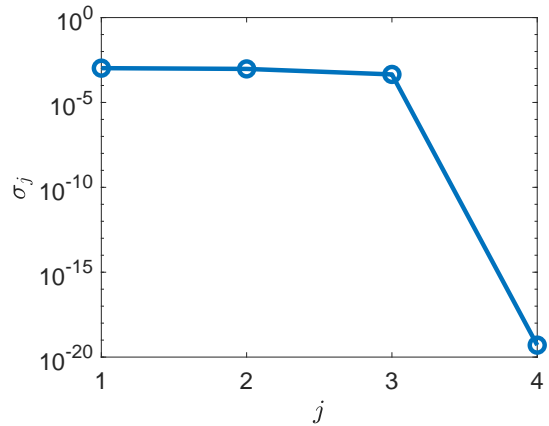}
    	\caption{Four singular values for data matrix $\mathbf{X}_1$ from the TDHF model.}
    	\label{fig:sing_HF}
    \end{figure}

\begin{figure}[htbp]
    	\centering
    	\includegraphics[width=0.45\textwidth]{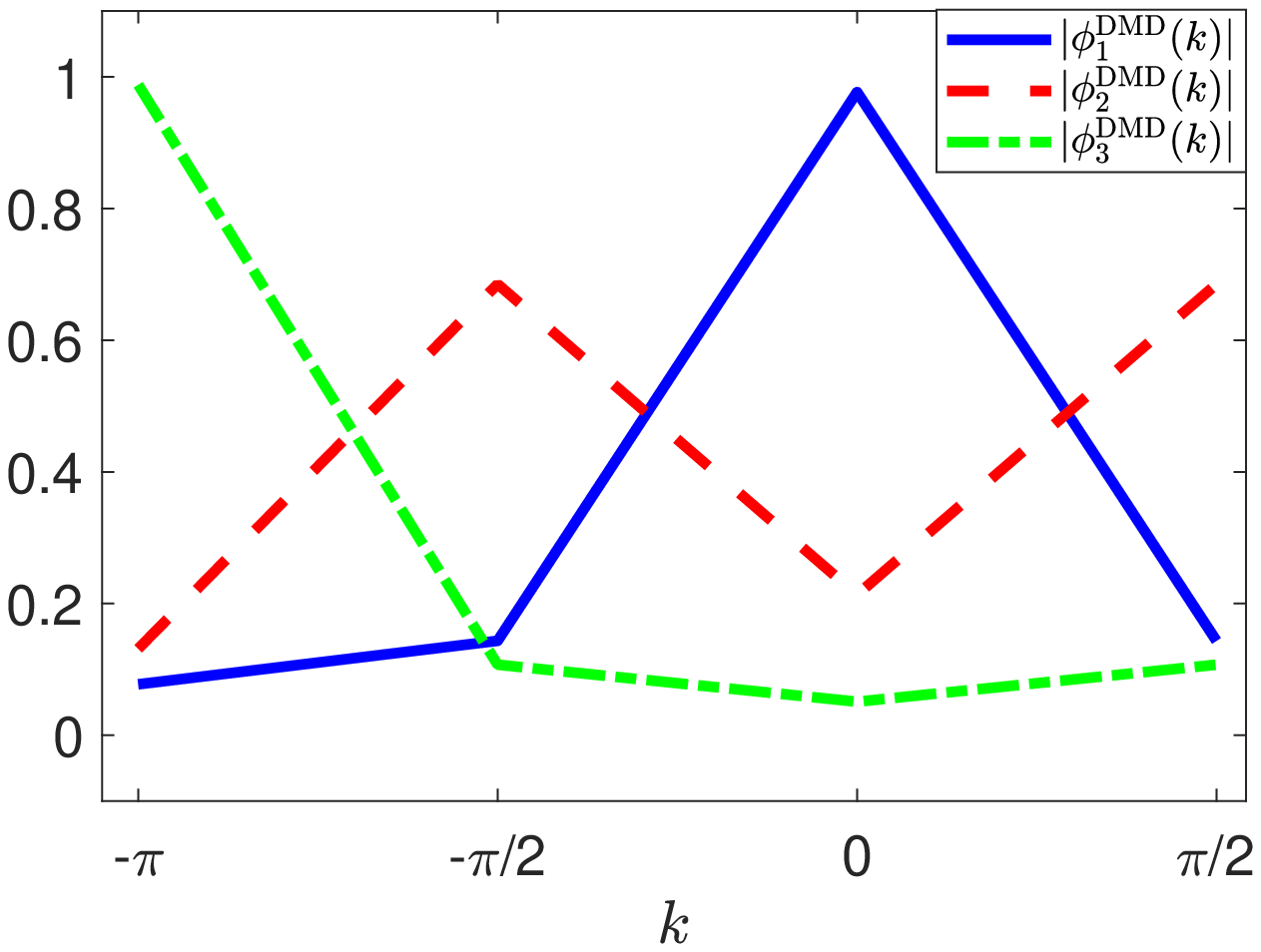}
    	\includegraphics[width=0.45\textwidth]{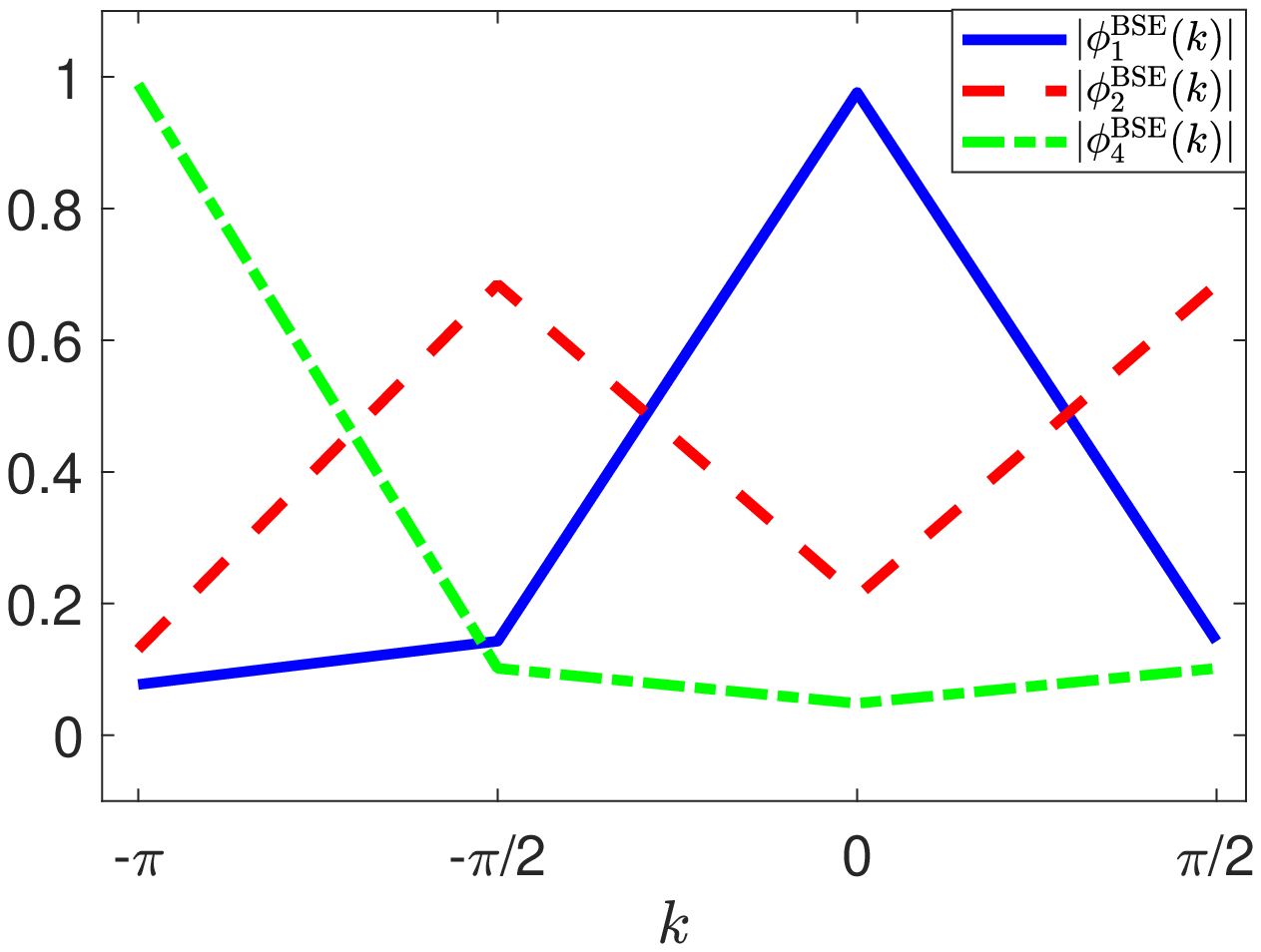}
    	\caption{\textit{left:} Three DMD modes obtained from snapshots of the numerical solution of the KBE for the 2-band model problem with HF self-energy. \textit{right:} Three eigenvectors of the corresponding BSE.}
    	\label{fig:comp_TDHF}
    \end{figure}

We took the first $m=500$ out of a total of $N=5000$ snapshots to perform the DMD of $\rho$. The nonzero singular values of the snapshot matrix $\mathbf{X}_1$ are plotted in Figure~\ref{fig:sing_HF}.
We can clearly see that the first three singular values are orders of magnitudes larger than the last singular value in this case. Consequently, we can approximate $\mathbf{X}_1$ by the three leading singular values and vectors. This approximation also yields three DMD modes obtained from \eqref{eq:dmdmodes}, which we plot in the left panel of Figure~\ref{fig:comp_TDHF}. The frequencies associated with these DMD modes, which are obtained from the eigenvalues of the matrix $\tilde{\mathbf{A}}$ defined in \eqref{eq:projA} are 
    \begin{equation}
        \omega_1^{\rm{DMD}} = -0.656-0.008i, \quad \omega_2^{\rm{DMD}} = -2.525-0.008i, \quad \omega_3^{\rm{DMD}} = -4.803-0.007i.
        \label{eq:TDHF_DMD_freq}
    \end{equation}
Due to the convention we used in \eqref{eq:xtexpand}, the real part of $\omega_j^{\rm{DMD}}$ corresponds to the frequency of temporal oscillation, and the imaginary part represents the rate of exponential growth or decay of the oscillation in time. In this case, the imaginary part of $\omega_j^{\rm{DMD}}$ should be zero. The  small imaginary components in \eqref{eq:TDHF_DMD_freq} are introduced by the numerical error in the approximate solution to the TDHF equation. %

It is well known that for TDHF, the absorption energy of the 2-band system and the corresponding exciton wavefunction can be obtained by performing a linear response 
analysis of the TDHF equation and solving the corresponding Bethe-Salpeter (or Casida) equation, which is an eigenvalue problem. The right panel of Figure~\ref{fig:comp_TDHF} shows the magnitudes of three eigenvectors of BSE match well with those of the DMD modes shown in the left panel. The corresponding eigenvalues are
	\begin{equation}
		\alpha_1 = 0.657, \quad \alpha_2 = 2.529, \quad \alpha_4 = 4.814,
	\end{equation}
which match well with the DMD frequencies listed in \eqref{eq:TDHF_DMD_freq}. The small difference between $\alpha_i$ and 
$-\rm{real}(\omega_i^{\mathrm{DMD}})$ is again due to the small numerical error present in the Runge-Kutta approximate solution of the TDHF equation. The excellent agreement suggests that the DMD modes are physical, and they properly describe the underlying exciton dynamics defined by the TDHF equation. 

\subsection{KBE with second Born approximation of the self-energy}
\begin{figure}[htbp]
    	\centering
        \begin{subfigure}[b]{0.45\textwidth}
  	       \includegraphics[width=1\textwidth]{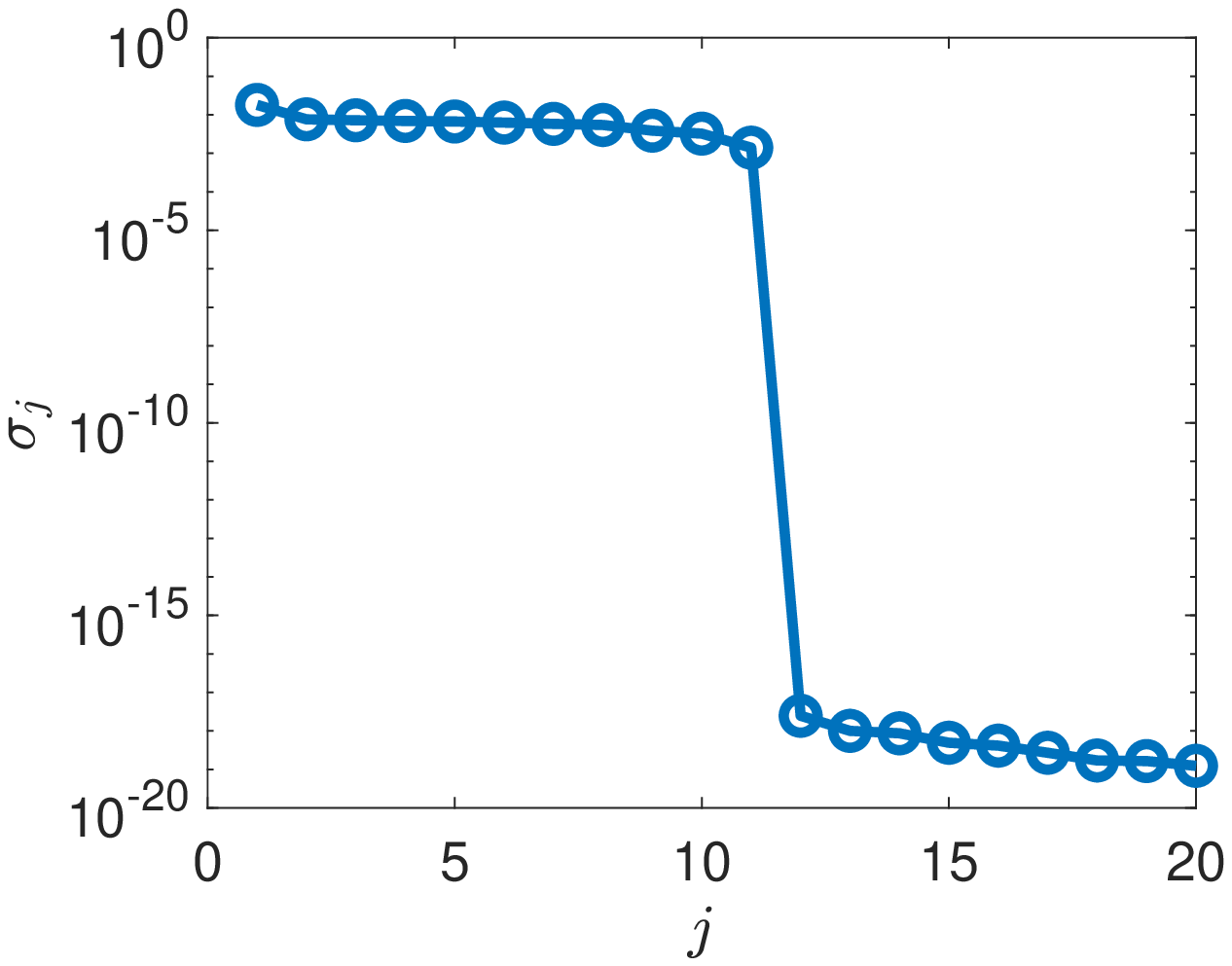}
           \caption{$I=0.001$}
    	\end{subfigure}
    	\begin{subfigure}[b]{0.45\textwidth}
    	    \includegraphics[width=1\textwidth]{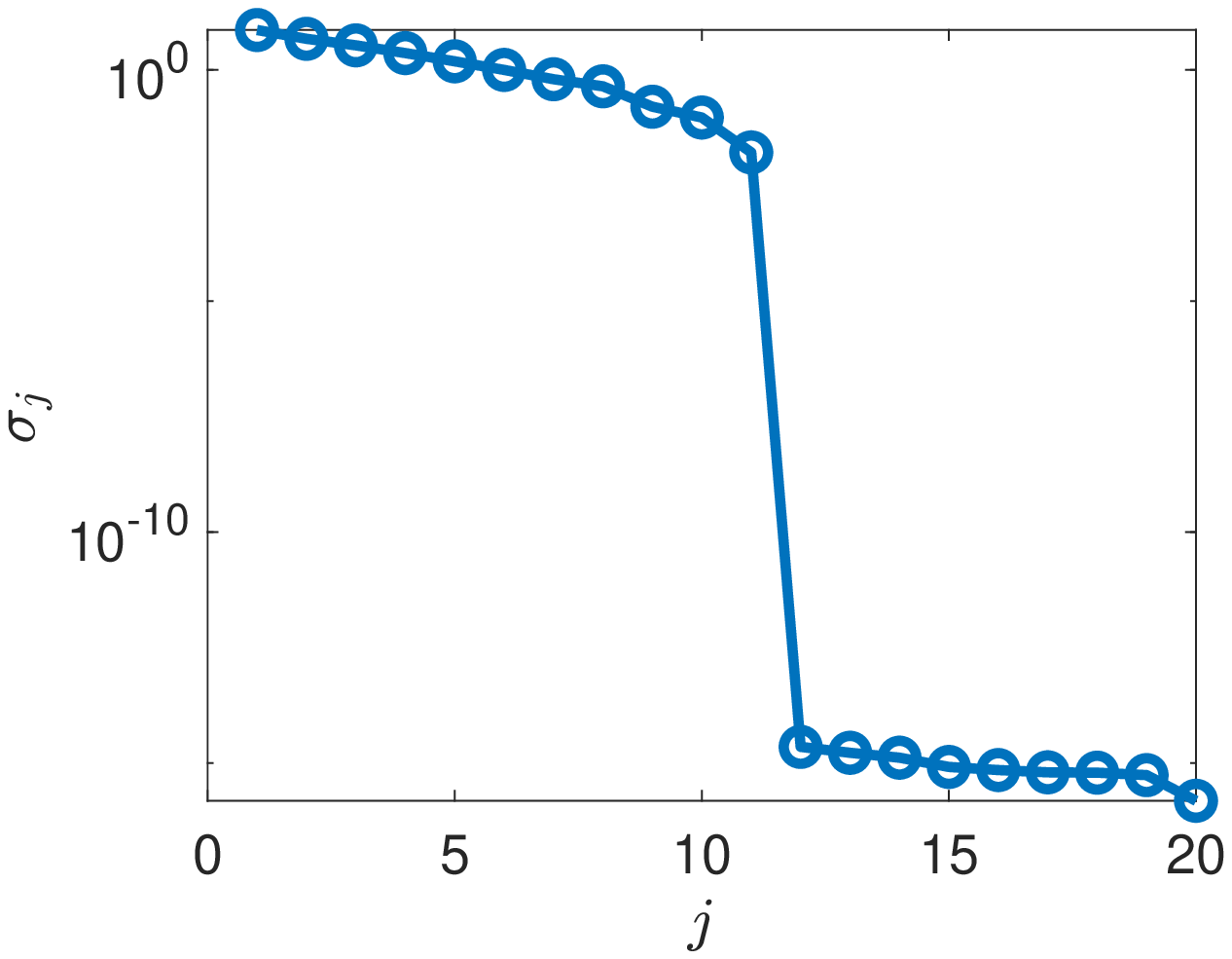}
    	    \caption{$I=1.5$}
        \end{subfigure}
        \caption{Singular values of the snapshot matrix $\mathbf{X}_1$ from the 2B model with different pulse intensities.}
        \label{fig:sing_2B}
    \end{figure}

In this example, the second Born approximation is used to construct the self-energy in the KBE \eqref{eq:KBE}. In addition, we assume that the system is driven by an instantaneous pulse $E(t) = I\delta(t)$ with $I$ being the field amplitude.
We sample the Brillouin zone with
$n=20$ $k$-points. For testing purpose, we solve the KBE for the time interval $[0,201]$ with time step $\Delta t=0.1$. We experimented with pulse amplitudes $I = 0.001$, $0.5$ and $1.5$ energy intensity. In each case, 
we used the first $m=500$ out of total $N=2010$ snapshots to perform the DMD analysis. The singular values of the snapshot matrix $\mathbf{X}_1$ in \eqref{eq:mats} for the weak ($I=0.001$) and 
strong ($I=1.5$) pulses are plotted in Figure \ref{fig:sing_2B}.  We can clearly see that, in both cases, the leading $11$ singular values are orders of magnitude larger than other singular values. This is also the case for the $\mathbf{X}_1$ generated from $I=0.5$.  Consequently, in all cases, we can approximate $\mathbf{X}_1$ by the leading 11 singular values and vectors.  

\begin{figure}[htbp]
    	\centering
        \includegraphics[width=0.45\textwidth]{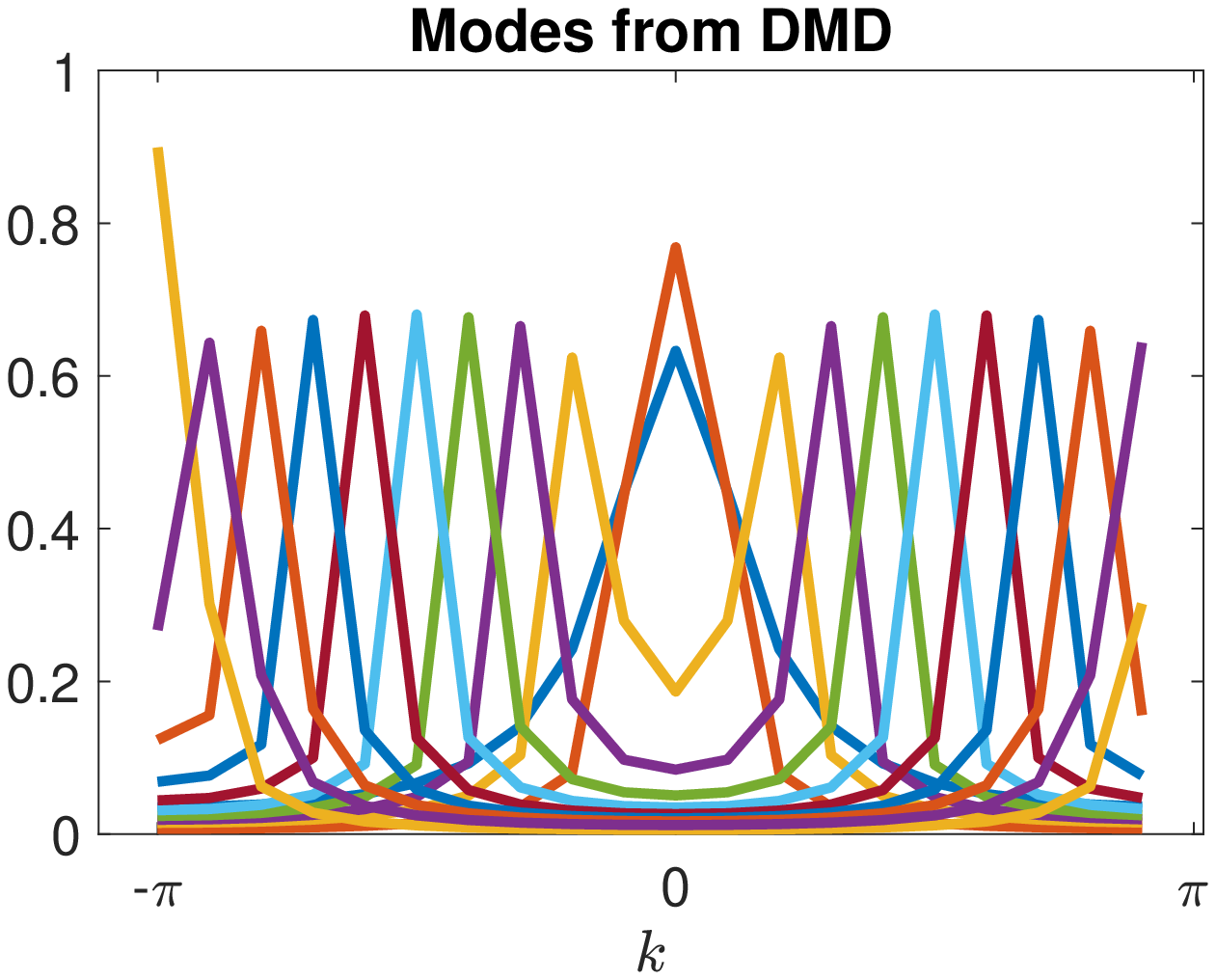}
        \includegraphics[width=0.45\textwidth]{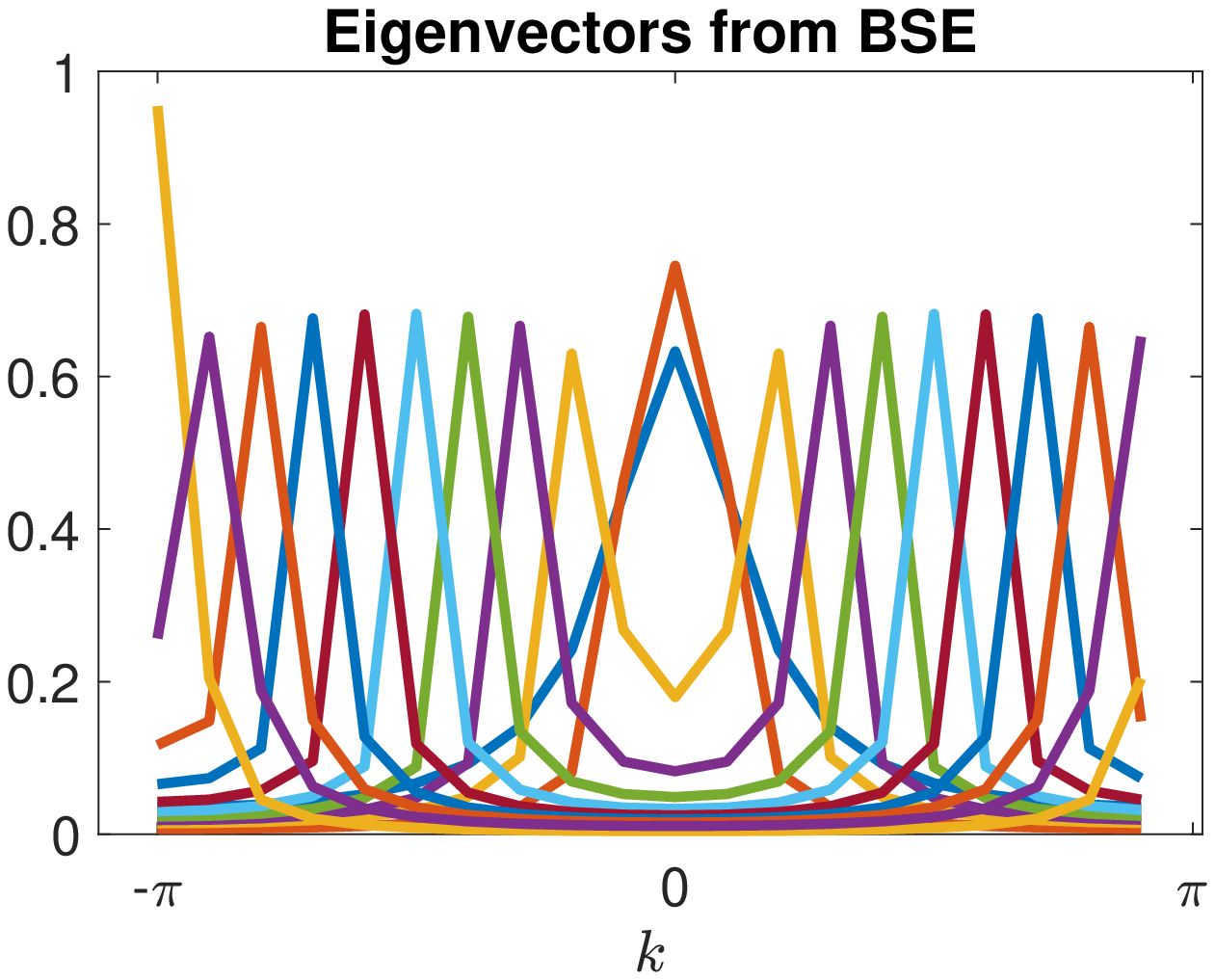}
        \caption{A comparison between the DMD modes obtained from snapshots of the solution to the KBE for the 2B model with $I=0.001$ (left) and the eigenvectors of the BSE Hamiltonian (right).} \label{fig:DMDvsBSE}
    \end{figure}

When the intensity of the pulse is small, we can perform a linear response analysis of $\rho(\cdot, t)$ by  solving a $n \times n$ BSE eigenvalue problem. Figure~\ref{fig:DMDvsBSE} shows that the eigenvectors of the BSE Hamiltonian (with Tamm-Dancoff approximation) associated with 11 smallest eigenvalues match well with the DMD modes computed from \eqref{eq:dmdmodes}.  The eigenvalues of the BSE Hamiltonian also match well with the real part of $\omega_j^{\mathrm{DMD}}$ for $j=1,2,...,11$.

%

\subsection{Comparison with Fourier Analysis}
In this section, we compare DMD analysis with Fourier spectral analysis, which 
has traditionally been used to identify key features of a one-dimensional trajectory $x(t)$. In such an analysis, we perform a discrete Fourier transform (DFT) of a uniformly sampled trajectory $\{x(t_1), x(t_2), ..., x(t_{N})\}$ to obtain
\[
f(\omega_\ell) =  \sum_{j=1}^{N}x(t_j)e^{-i \omega_\ell (t_j-t_1)}, 
\]
where 
\begin{equation}
	\omega_\ell =  2\pi (\ell-1)/(N\Delta t) + 2z\pi, \quad \ell = 1, 2, ..., N, \; \forall z\in\mathbb{Z}.
\end{equation} 
Note that the $2z\pi$ term is included above to match some frequencies computed from DMD that are not within the same period. Clearly, the larger the magnitude of $f(\omega_\ell)$, the more important $\omega_\ell$ is for describing the dynamics exhibited by $x(t)$. If $x(t)$ can be characterized by a few frequencies,
$|f(\omega_\ell)|$ will exhibit a few peaks.  

\begin{figure}[htbp]
    	\centering
        \includegraphics[width=0.6\textwidth]{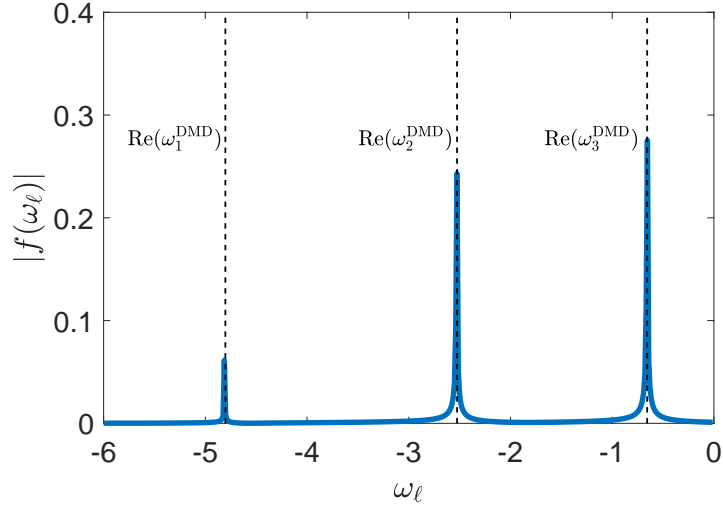}
        \caption{The magnitude of $f(\omega_\ell)$ obtained from the DFT of the polarizability associated with the TDHF simulation of a 2-band model. The real part of the DMD frequencies are marked by dotted lines.}
        \label{fig:tdhf_dft_freq}
    \end{figure}

\begin{figure}[htbp]
    	\centering
        \includegraphics[width=0.6\textwidth]{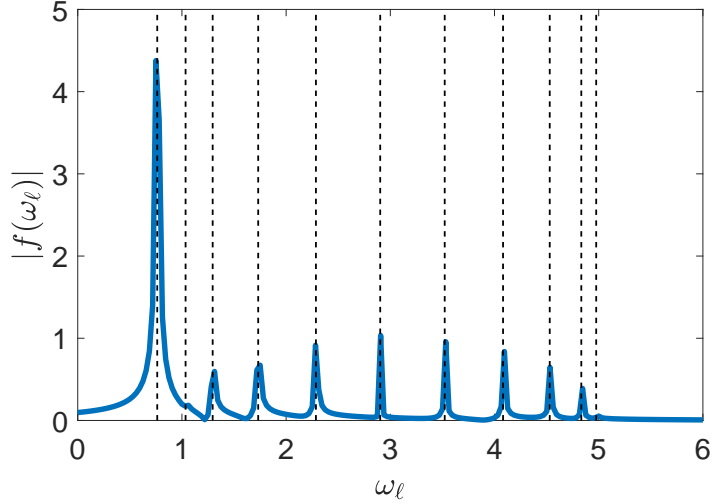}
        \caption{The magnitude of $f(\omega_\ell)$ obtained from the DFT of the polarizability associated with the solution of the KBE with a second Born approximation to the self energy and driven by an instantaneous field with intensity $I = 0.001$. The real part of the DMD frequencies are marked by dotted lines.}
        \label{fig:2Born_dft_freq}
    \end{figure}

In Figure~\ref{fig:tdhf_dft_freq}, we plot the magnitude of $f(\omega_\ell)$ 
obtained from performing a DFT of the polarizability $P(t)$ defined as
\begin{equation}
P(t)=\sum_{\mathbf{k}}\text{Tr}(\rho_\mathbf{k}(t)\hat{d_\mathbf{k}}),
\label{eq:polar}
\end{equation}
where $\mathbf{k}$ denotes a $k$-point, and $\hat{d_\mathbf{k}}$ is a dipole matrix.
In our two band model with a constant dipole matrix element approximation, the polarization is simply the sum of the two off-diagonal elements of the density matrix. By analyzing the polarization in the linear response regime, we can get the exciton energy and wavefunctions.
The positions of the three peaks of $f(\omega_\ell)$ in Figure~\ref{fig:tdhf_dft_freq} are \begin{equation*}
        \omega_1^{\rm{DFT}} = -0.654, \quad \omega_2^{\rm{DFT}} = -2.526, \quad \omega_3^{\rm{DFT}} = -4.813.
    \end{equation*}
    These frequencies match well with the real parts of the three DMD frequencies shown in~\eqref{eq:TDHF_DMD_freq}.
When the Fourier analysis is applied to the polarizability obtained from the  numerical solution of the KBE with a second Born approximation to the self-energy, we observe a similar match between the peak positions of $|f(\omega_\ell)|$ and  the real part of $\omega_j^{\rm{DMD}}$  obtained from \eqref{eq:dmdev} and \eqref{eq:dmdfreq} as long as the pulse intensity $I$ 
is relatively small. This can be clearly seen in Figure~\ref{fig:2Born_dft_freq} 
in which the magnitude of $f(\omega_\ell)$ is plotted for $I=0.001$. The positions of the peaks match well with the real part of  the DMD frequencies which are marked by vertical dotted lines.

In addition to frequencies, the DMD analysis also provides spatial (momentum) modes $\Phi$
associated with different frequencies.  Similar type of modes, which we will call 
Fourier modes, can be obtained from the Fourier analysis, although the procedure for  finding these modes is a bit cumbersome. 

Instead of performing a DFT to the polarizability, we can perform a set of DFTs to $\rho(k_s,t)$ for each $k$-point to obtain $f_s(\omega_\ell^{\mathrm{DFT}})$, where $\ell = 1,2,...,\hat{r}$ if we have $\hat{r}$ DFT frequencies, and $s = 1, ..., n$. The vector 
\begin{equation}\label{eq:dft_mode}
    \phi_\ell^{\mathrm{DFT}} = \frac{[f_1(\omega_\ell^{\mathrm{DFT}}), f_2(\omega_\ell^{\mathrm{DFT}}), ... f_n(\omega_\ell^{\mathrm{DFT}})]^T}{\|[f_1(\omega_\ell^{\mathrm{DFT}}), f_2(\omega_\ell^{\mathrm{DFT}}), ... f_n(\omega_\ell^{\mathrm{DFT}})]\|_{l^2}}, \quad \ell=1, ..., \hat{r}
\end{equation}
defines
the $\ell$th Fourier mode associated with the frequency $\omega_\ell^{\mathrm{DFT}}$.  The $\ell$th Fourier mode makes a significant contribution to $\rho(t)$ if $|f_{s}(\omega_\ell^{\mathrm{DFT}})|$ is sufficiently large for some $s$.  

\begin{figure}[htbp]
    	\centering
    	\includegraphics[width=0.47\textwidth]{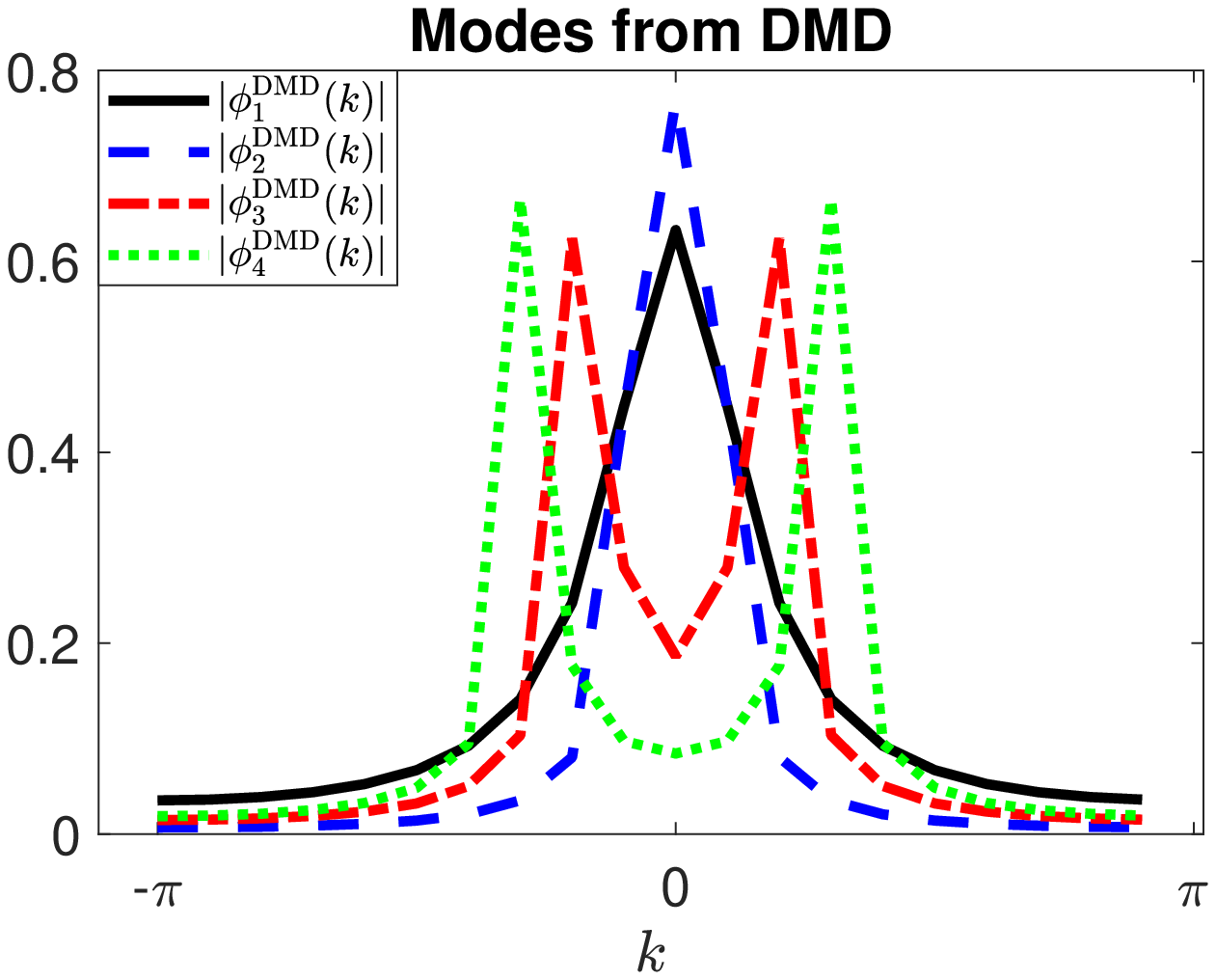}
    	\includegraphics[width=0.47\textwidth]{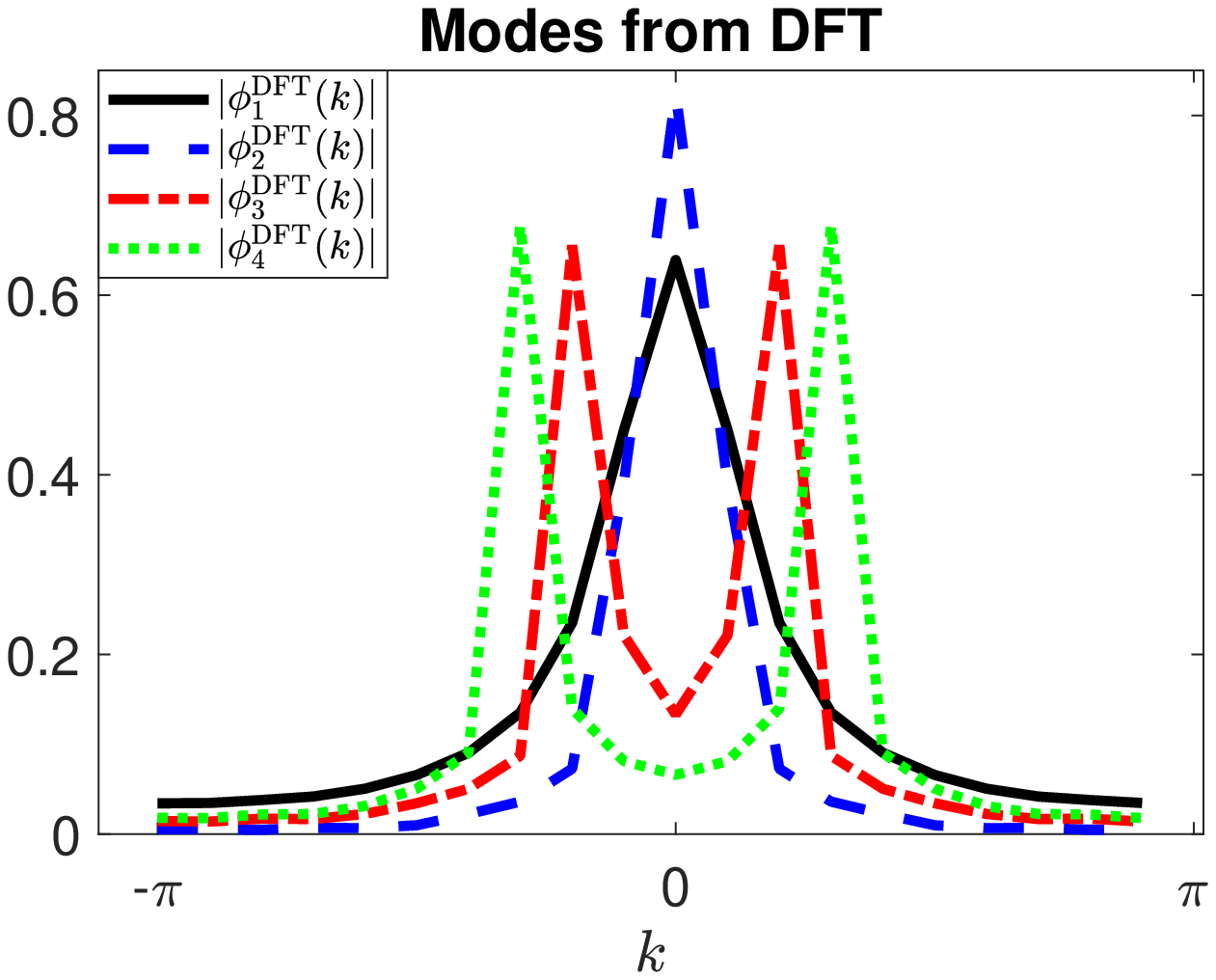}
    	\caption{DMD modes (left) match well with the Fourier modes (right) with matching frequencies for the two-band model driven by a field with intensity $I = 0.001$.}
    	\label{fig:comp_2B0p001}
    \end{figure}

Figure~\ref{fig:comp_2B0p001} shows that the first four Fourier modes obtained from the 
solution of the KBE with a second Born self-energy approximation and driven by a pulse with intensity $I=0.001$ match well with the corresponding DMD modes after being scaled by a phase factor. (i.e. like eigenvectors,
DMD modes are unique up to a phase scaling factor $e^{i\theta}$ for some phase angle $\theta$.)

\begin{figure}[htbp]
    	\centering
        \includegraphics[width=0.6\textwidth]{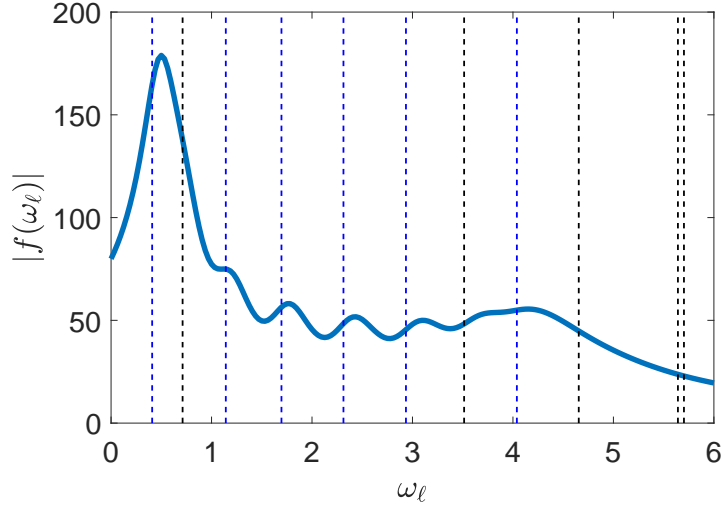}
        \caption{The magnitude of $f(\omega_\ell)$ obtained from the DFT of the polarizability for the two-band model driven by a field with intensity $I=1.5$. The DMD frequencies are marked by dotted lines. The DMD dotted lines corresponding to matching frequencies are colored in blue.} 
        \label{fig:diff_DMD_DFT_freq}
    \end{figure}

However, when the pulse intensity increases to $I=0.5$ and $I=1.5$,  not all frequencies identified from the Fourier analysis can be matched with those obtained from the DMD analysis.  In fact, $f(\omega_\ell)$ cannot be characterized by a few isolated peaks. Figure~\ref{fig:diff_DMD_DFT_freq} shows that when $I=1.5$, we can only match six frequencies obtained from the Fourier analysis with those obtained from DMD (marked by blue dotted lines). These matches are not perfect.

\begin{figure}[htbp]
    	\centering
    	\includegraphics[width=0.47\textwidth]{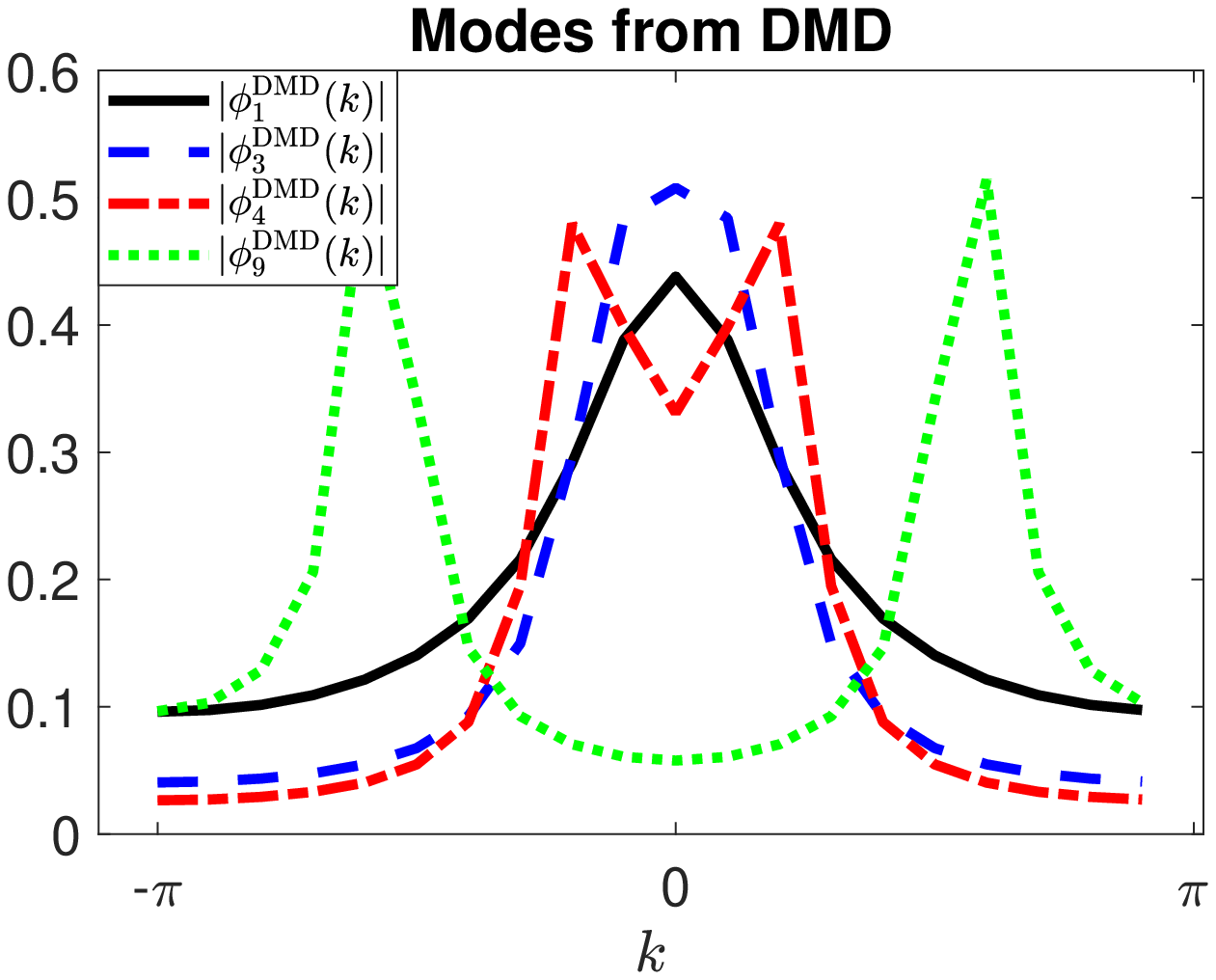}
    	\includegraphics[width=0.47\textwidth]{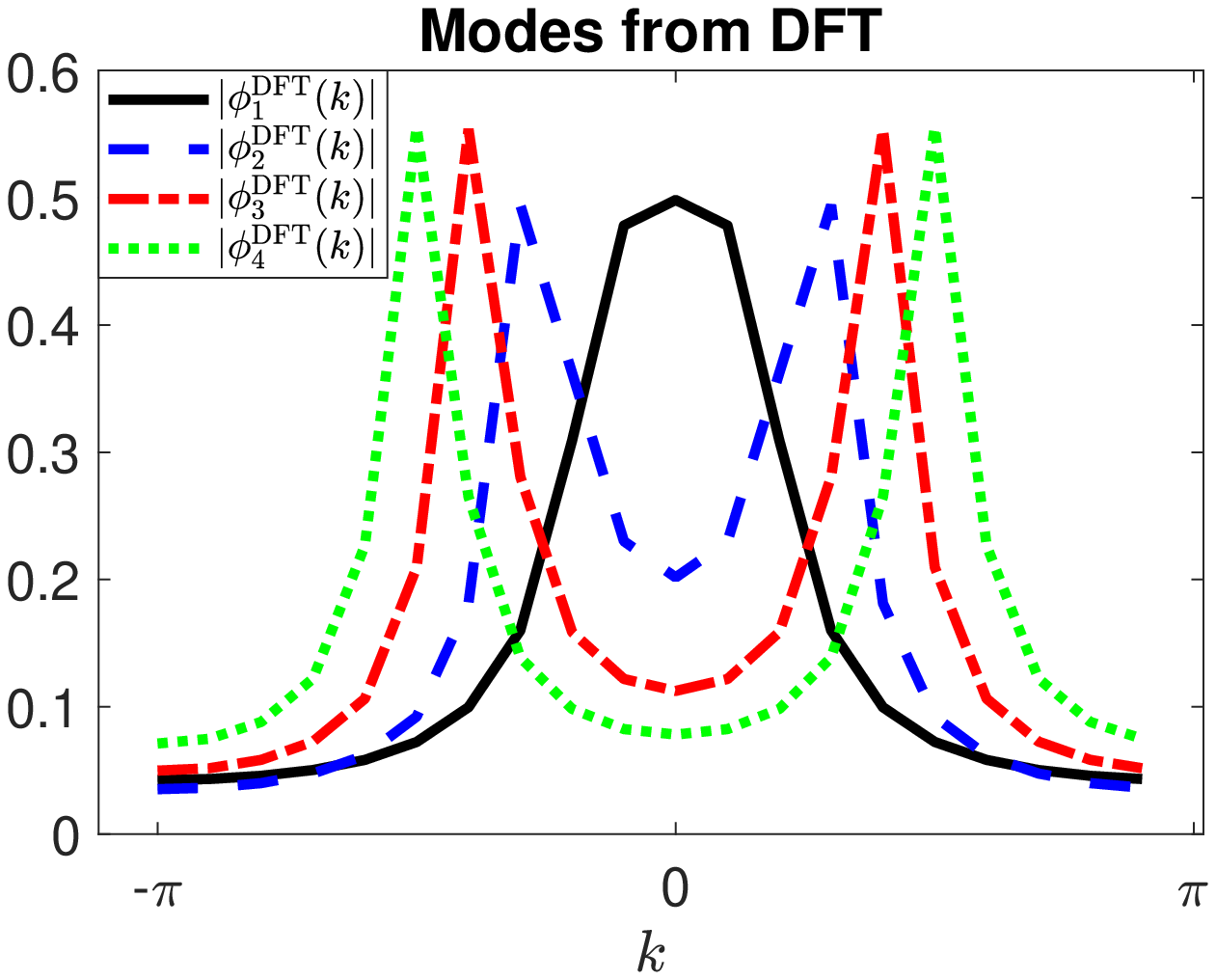}
    	\caption{The four most significant modes from DMD (left) and from DFT (right) for the 2-band model driven by a field with intensity $I = 1.5$. 
    	}
    	\label{fig:comp_2B1p5}
    \end{figure}

In Figure~\ref{fig:comp_2B1p5}, we compare 4 most significant modes obtained from DMD and DFT. 
The significance of each DFT mode can be quantified by the height of the peak associated with the frequency of that mode.  The significance of each DMD mode can be measured by the magnitude of the corresponding expansion coefficient in the reconstruction or extrapolation (see the next section) of the trajectory as described by \eqref{eq:evol_DMD}. 

The frequencies of the four most significant DMD modes are
    \begin{align*}
    &\omega_1^{\rm{DMD}} = -0.581+0.226i, \quad \omega_3^{\rm{DMD}} = 0.409+0.030i, \\
    &\omega_4^{\rm{DMD}} = 0.712+0.029i, \quad \omega_9^{\rm{DMD}} = 2.935+0.049i.
    \end{align*}
The real parts of these frequencies which describe the oscillatory behavior of $\rho(t)$ in real time by our convention do not closely match (with the exception of $\omega_3^{\rm{DMD}}$) the four most significant frequencies obtained from the DFT, which are 
\begin{equation*}
    \omega_1^{\rm{DFT}} = 0.406, \quad \omega_2^{\rm{DFT}} = 1.125, \quad \omega_3^{\rm{DFT}} = 1.782, \quad \omega_4^{\rm{DFT}} = 2.438.
    \end{equation*}
Note that the real part of $\omega_3^{\rm{DMD}}$ matches well with $\omega_1^{\rm{DFT}}$. For these two matching frequencies, the corresponding DMD and Fourier modes also match well as we can see in  Figure~\ref{fig:comp_2B1p5}. All the other three DMD modes are different from the other three Fourier modes. In particular, the most significant DMD mode $\phi_1^{\rm{DMD}}$ is not seen in the Fourier analysis. The DMD mode $\phi_9^{\rm{DMD}}$ looks similar to both $\phi_3^{\rm{DFT}}$ and $\phi_4^{\rm{DFT}}$. However, their values are clearly different at $k=-\pi$ and $k=0$.

\subsection{Extrapolation of the density dynamics}
The discrepancy between the frequencies identified by Fourier analysis and DMD analysis raises the question about which analysis is more useful or reliable. 
In the case of a low intensity driving field, we can compare the spectral modes with the eigenvectors of the BSE problems. However, when the pulse intensity is high, we can no longer rely on  the BSE which, is validate in the linear response regime, to validate the spectral modes.

One way to assess the validity or quality of the spectral modes is to examine how well they can be used to reconstruct the sampled density trajectory and extrapolate the density dynamics outside of the sampling window.


For DMD analysis, we follow \eqref{eq:evol_DMD} to reconstruct and extrapolate the density trajectory $\rho^{\mathrm{DMD}}$ at k-point $k_s$ as
\begin{equation}
\mathbf{\rho}^{\mathrm{DMD}}(k_s, t)\approx \sum_{\ell=1}^r\mathbf{\phi}_\ell^{\mathrm{DMD}}(k_s)\exp(i\omega_\ell^{\mathrm{DMD}} t)b_\ell, \quad s=1, ..., n,
\label{eq:dmd_recon}
\end{equation}
where $b_\ell$ are obtained either from the projection of initial density onto the DMD modes as \eqref{eq:b1}, or from performing a linear least squares fit on the snapshots used to perform the DMD analysis as given in \eqref{eq:b2}.

On the other hand, the Fourier analysis in the above subsection cannot be directly applied to do the extrapolation, as the whole trajectory instead of the sampled trajectory is required there.
In order to use the sampled snapshots only, we need to first pad them with zeros before taking the discrete Fourier transform, i.e., we construct $\{\tilde{\rho}(k_s,\cdot)\}$ as
\begin{equation}
	\tilde{\rho}(k_s, t_j) = \left\{ 
	    \begin{aligned}
		 & \rho(k_s, t_j), \quad j=1, 2, ..., m\\
		 & 0,  \qquad\quad j=m+1, ..., N
		\end{aligned}\right.
\end{equation}
for $s = 1, ..., n$, and take the discrete Fourier transform  of the polarization $P(t)$ of these trajectories. 

\begin{figure}[htbp] \centering
	\includegraphics[width=0.46\textwidth]{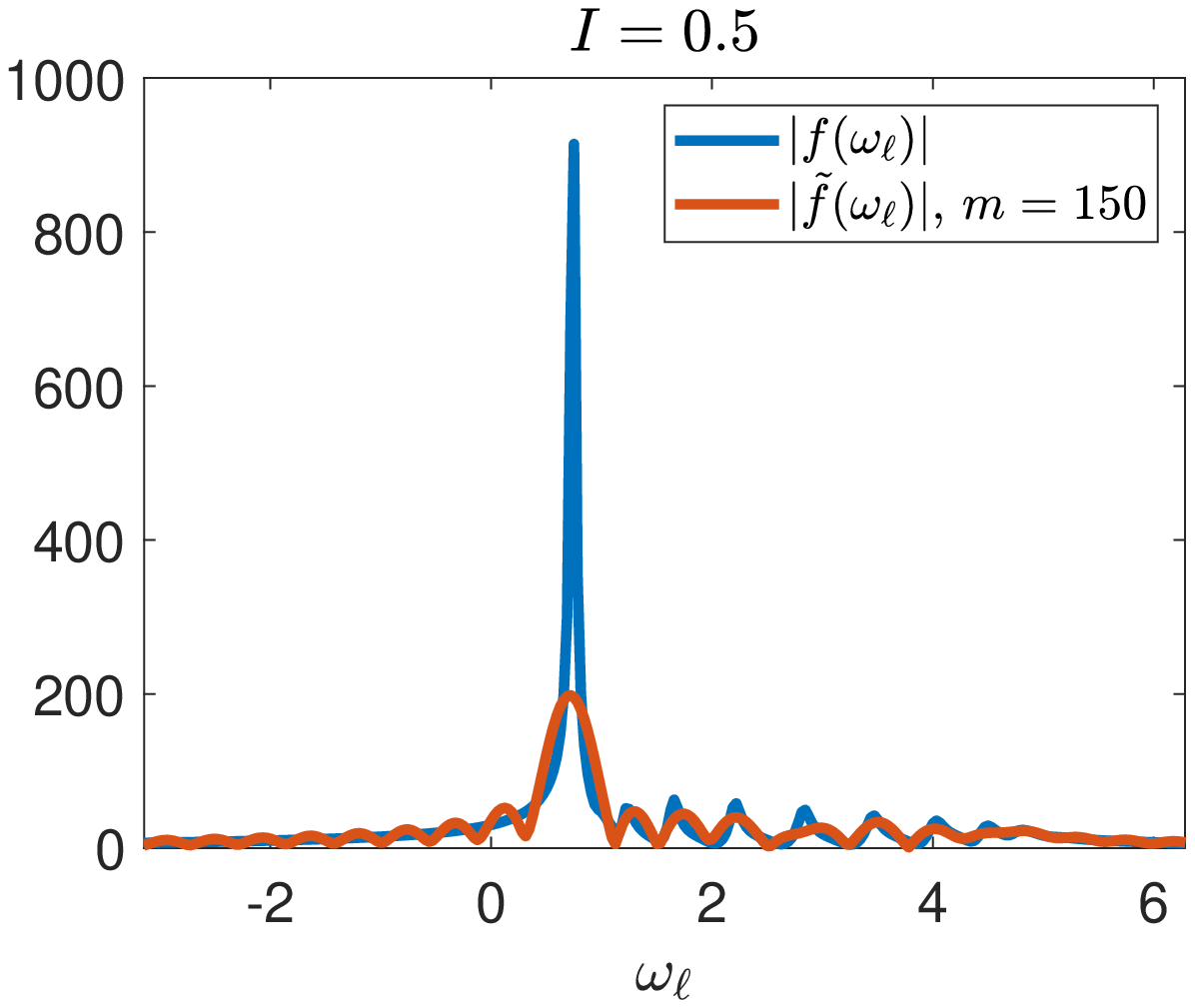}
	\includegraphics[width=0.46\textwidth]{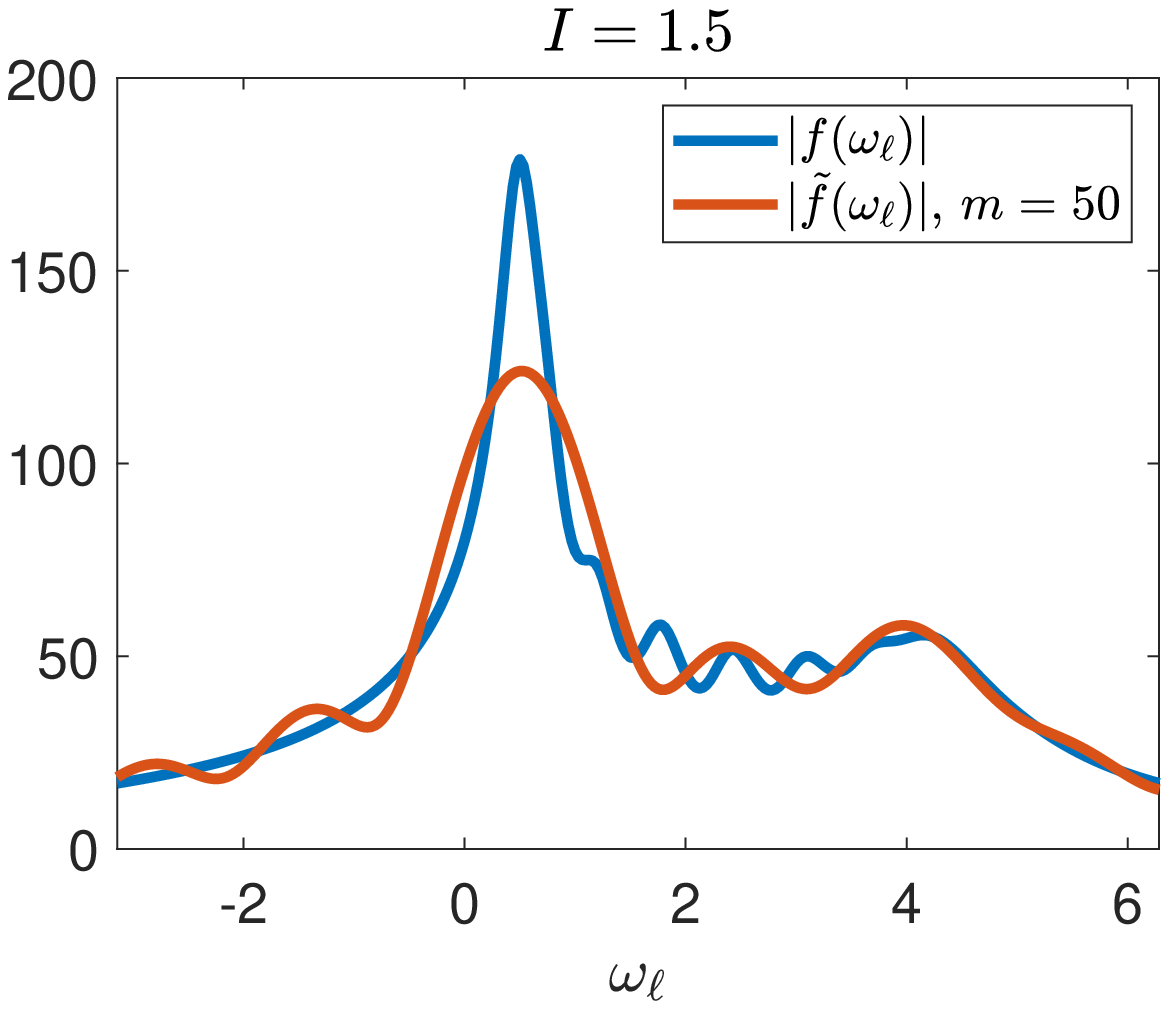}
	\caption{Comparison between the magnitude of $f(\omega_\ell)$ and $\tilde{f}(\omega_\ell)$ when $I=0.5$ and $I=1.5$.}
	\label{fig:dft_trun}
\end{figure}

This is equivalent to convolving the discrete Fourier transform of $P(t)$ \eqref{eq:polar} with a sinc function. Such a convolution broadens the high peaks in $f(\omega_\ell)$ and introduces artifical wiggles not present in $f(\omega_\ell)$ as can be seen
in Figure \ref{fig:dft_trun} where the discrete Fourier transforms of the full polarization trajectory  and the truncated and zero padded
polarization are compared for models with driving field intensities $I=0.5$ and $I=1.5$. We take $m=150$ and $m=50$ respectively for the two cases, in order to show the comparison clearly.

Once the frequency peaks of $\tilde{f}(\omega_\ell)$ are identified, we denote
them by $\tilde{\omega}_\ell^{\mathrm{DFT}}$, $\ell = 1,2,...,\tilde{r}$ to reconstruct an approximate trajectory as 
\begin{equation}
\mathbf{\rho}^{\mathrm{DFT}}(t)\approx \sum_{\ell=1}^{\tilde{r}}\tilde{\mathbf{\phi}}^{\mathrm{DFT}}_\ell\exp(i \tilde{\omega}_\ell^{\mathrm{DFT}} t)c_\ell\exp(d_lt),\quad d_l\leq 0,
\label{eq:dft_recon}
\end{equation}
where $\tilde{\mathbf{\phi}}^{\mathrm{DFT}}_\ell$ is the DFT mode associated with $\tilde{\omega}_{\ell}^{\mathrm{DFT}}$ as defined in \eqref{eq:dft_mode}, $c_\ell$ and $d_\ell$ are parameters to be determined, with $\exp(d_\ell t)$ being introduced to account for the exponential decay of the dynamics. We fit the model \eqref{eq:dft_recon} to the sampled snapshots in the least-squares sense, and obtain the fitting coefficients $\mathbf{c}:=(c_1, ..., c_{\tilde{r}})^T$ and $\mathbf{d}:=(d_1, ..., d_{\tilde{r}})^T$ by solving a 
nonlinear least squares problem.

To compare the reconstructed trajectories, we perform a renormalization so that
	\begin{equation}
		\|\rho^{\mathrm{a}}(k_s, t_1:t_m)\|_{l^2} = \|\rho(k_s, t_1:t_m)\|_{l^2}, \quad s=1, ..., n,
    \end{equation} 
where $\rm{a} = \text{DMD or DFT}$, and $\rho(k_s, t)$ is the sampled data. 

\begin{figure}[htbp]
	\centering
	\includegraphics[width=0.45\textwidth]{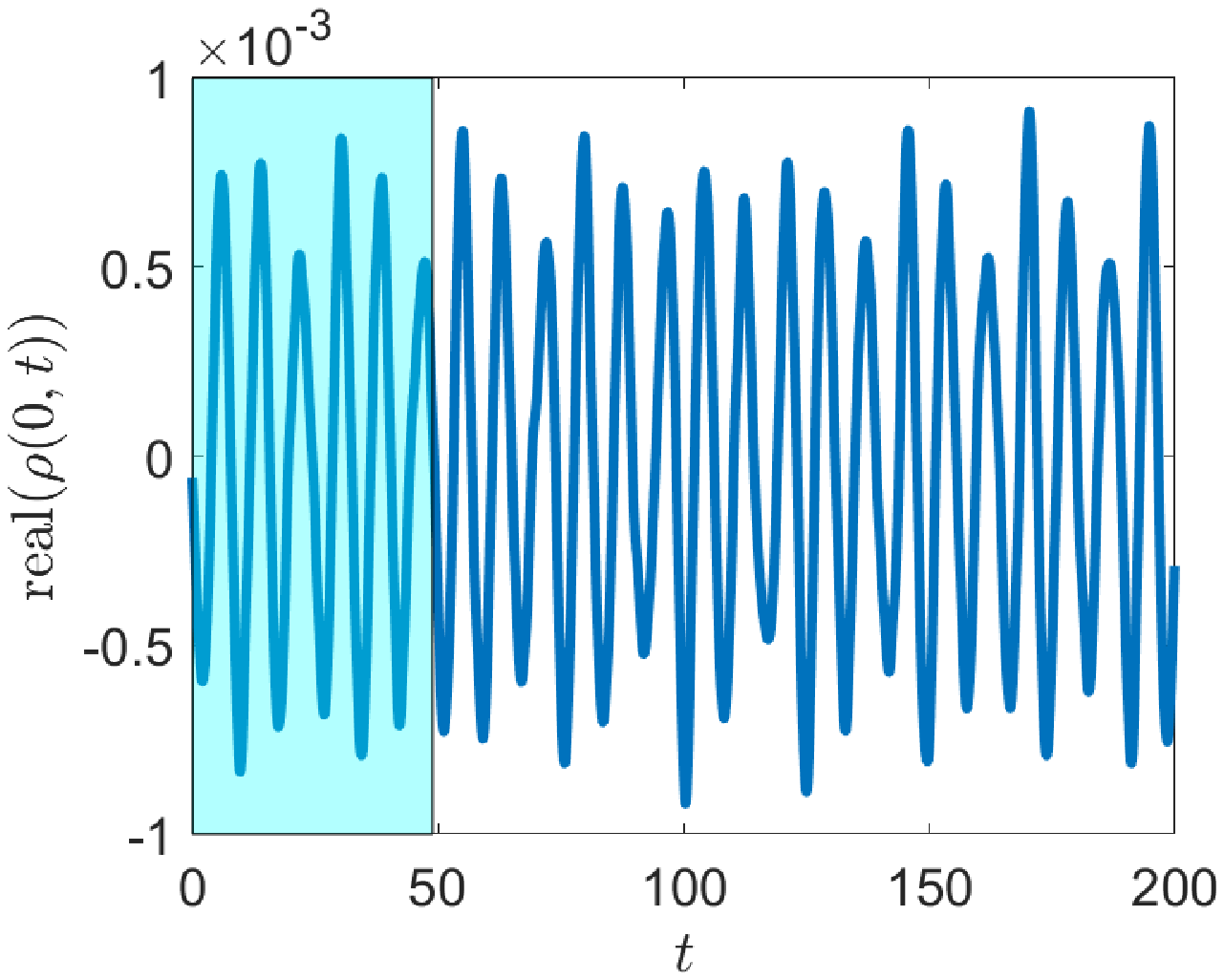}
	\includegraphics[width=0.45\textwidth]{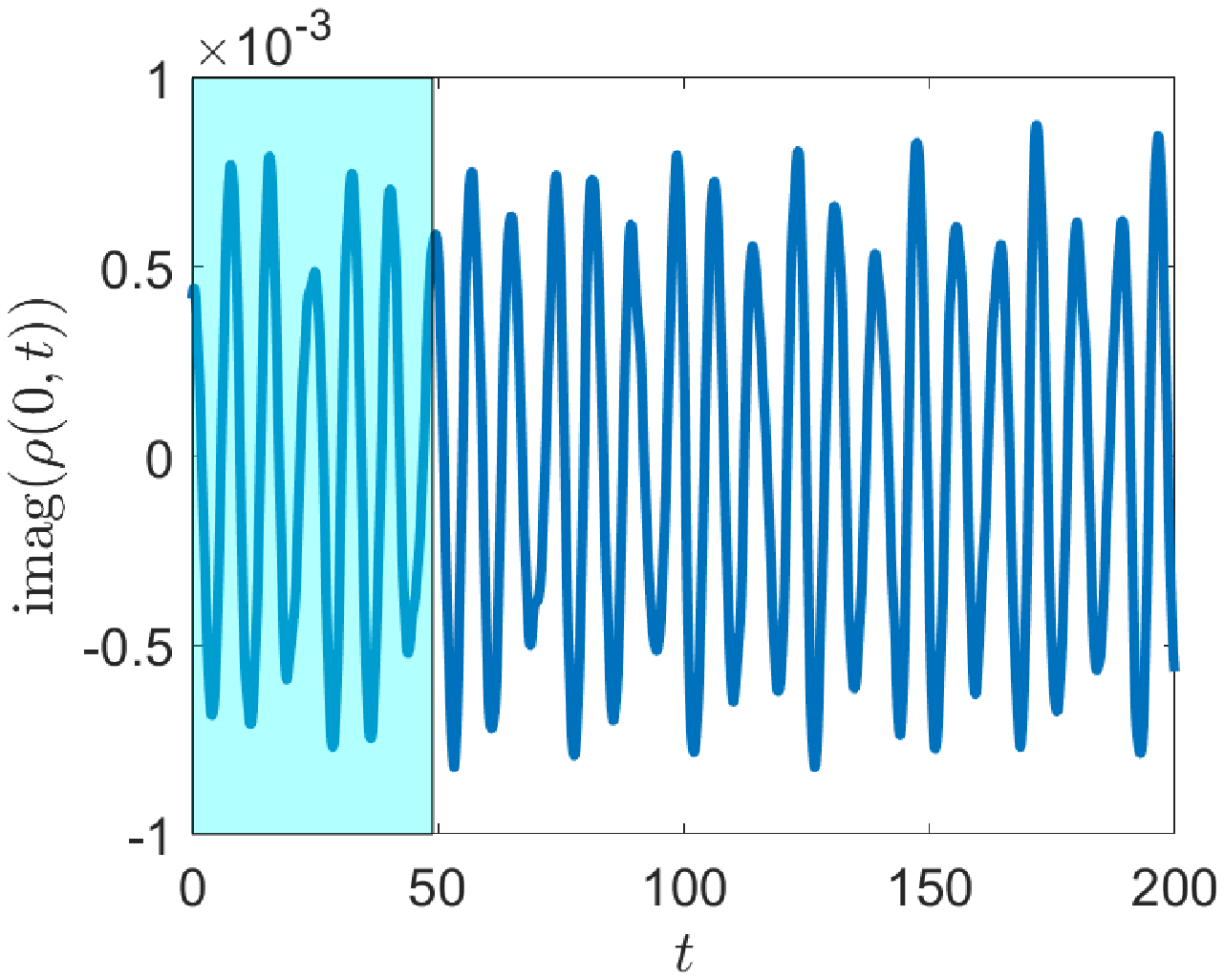}
	\includegraphics[width=0.45\textwidth]{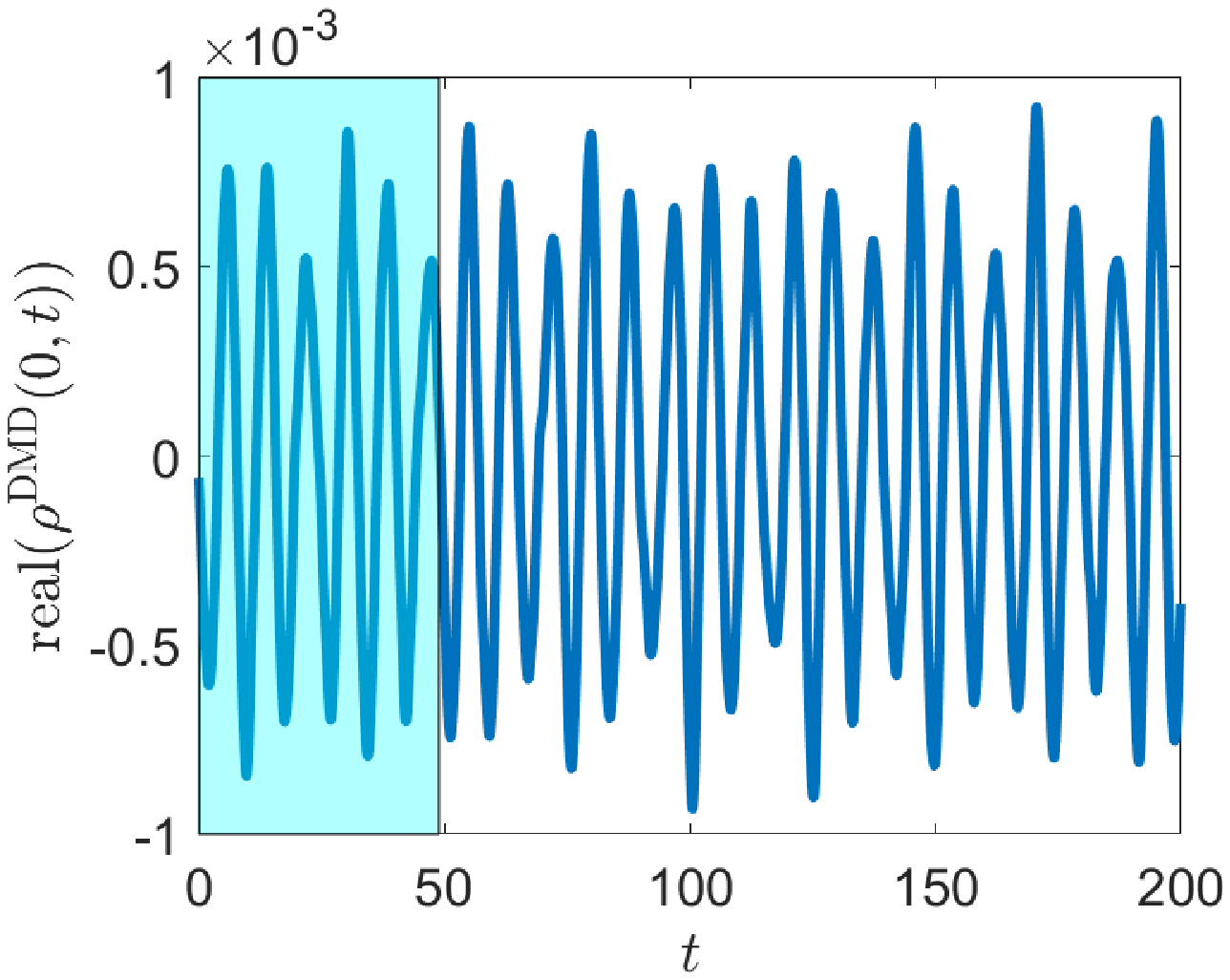}
	\includegraphics[width=0.45\textwidth]{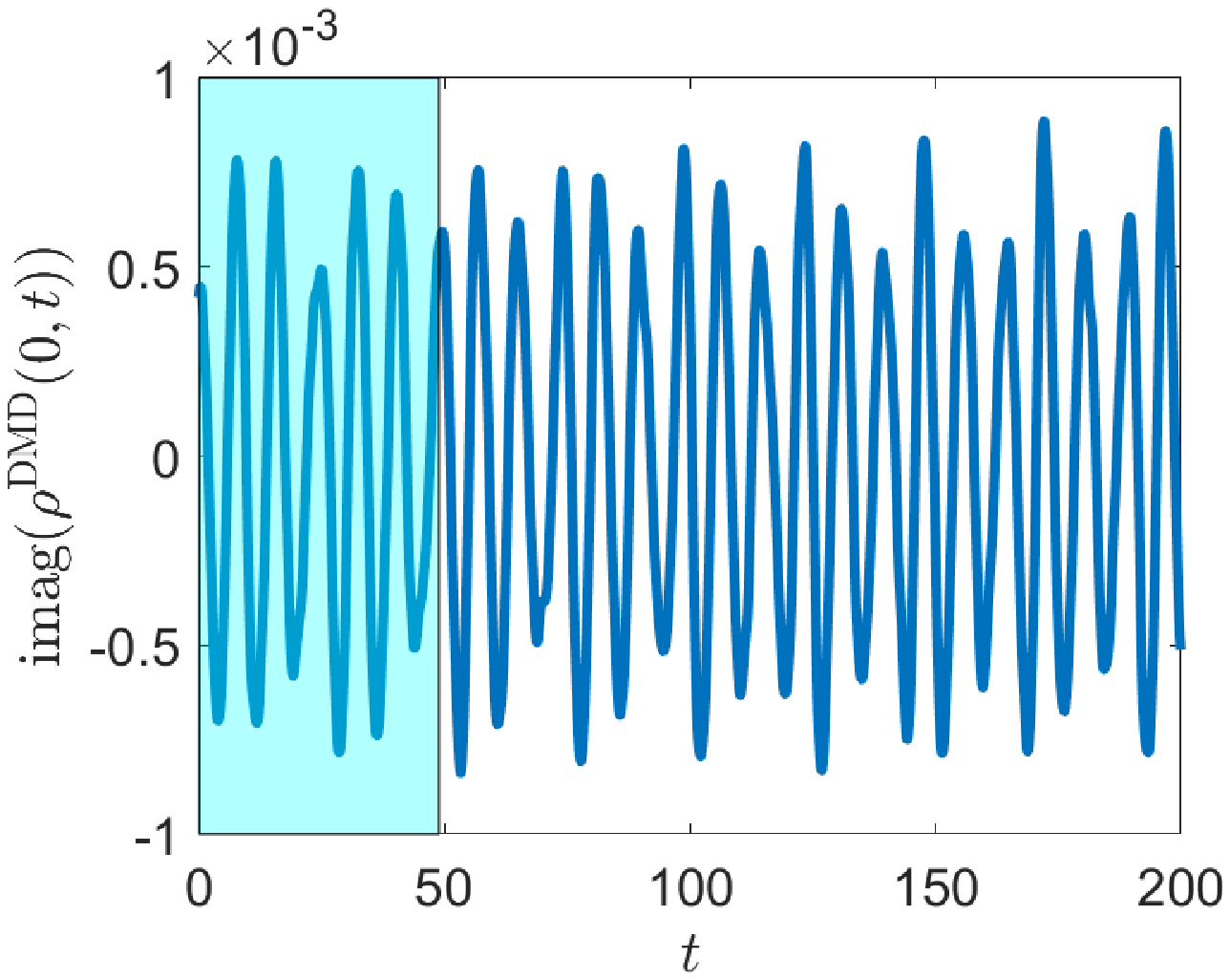}
    \includegraphics[width=0.45\textwidth]{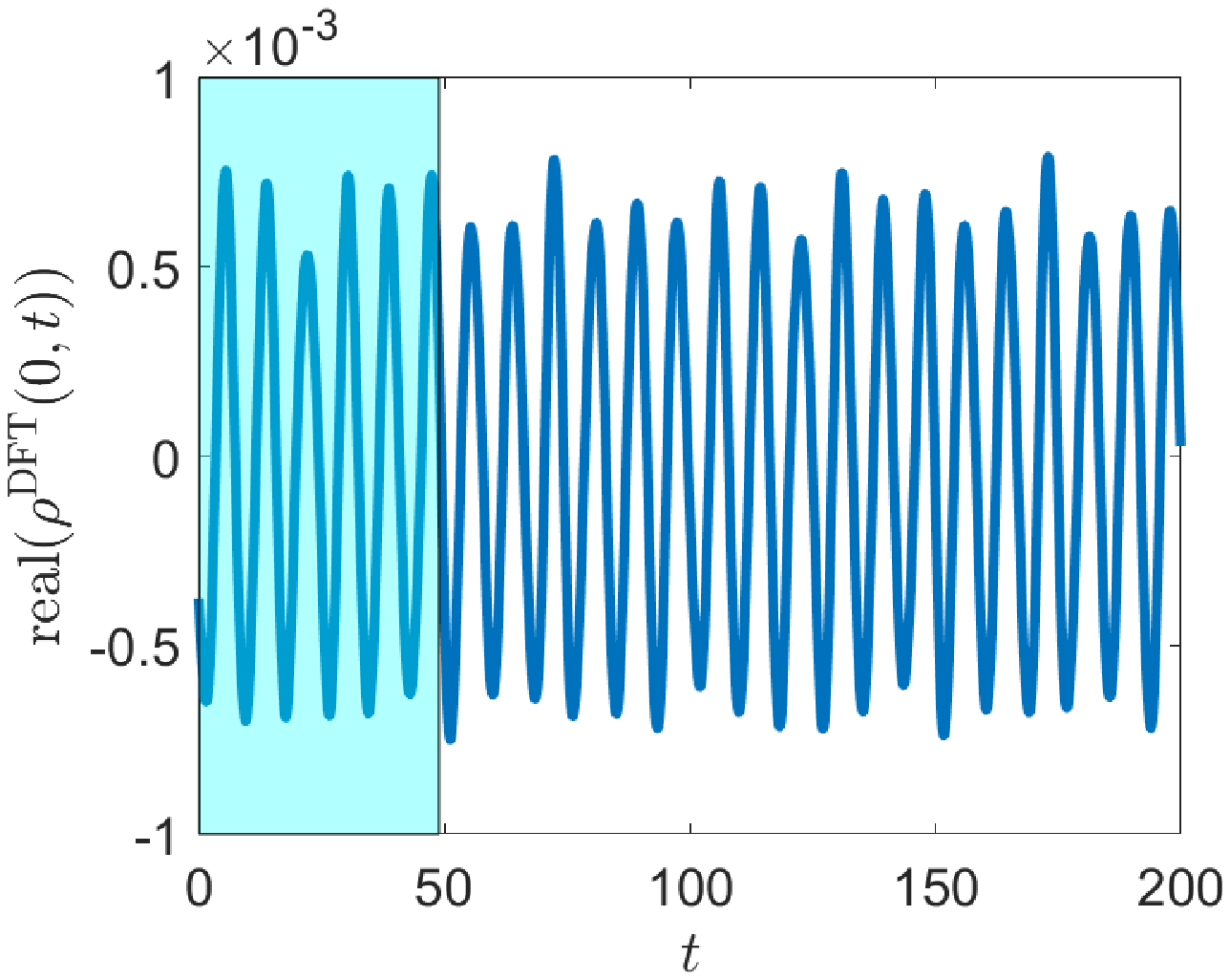}
    \includegraphics[width=0.45\textwidth]{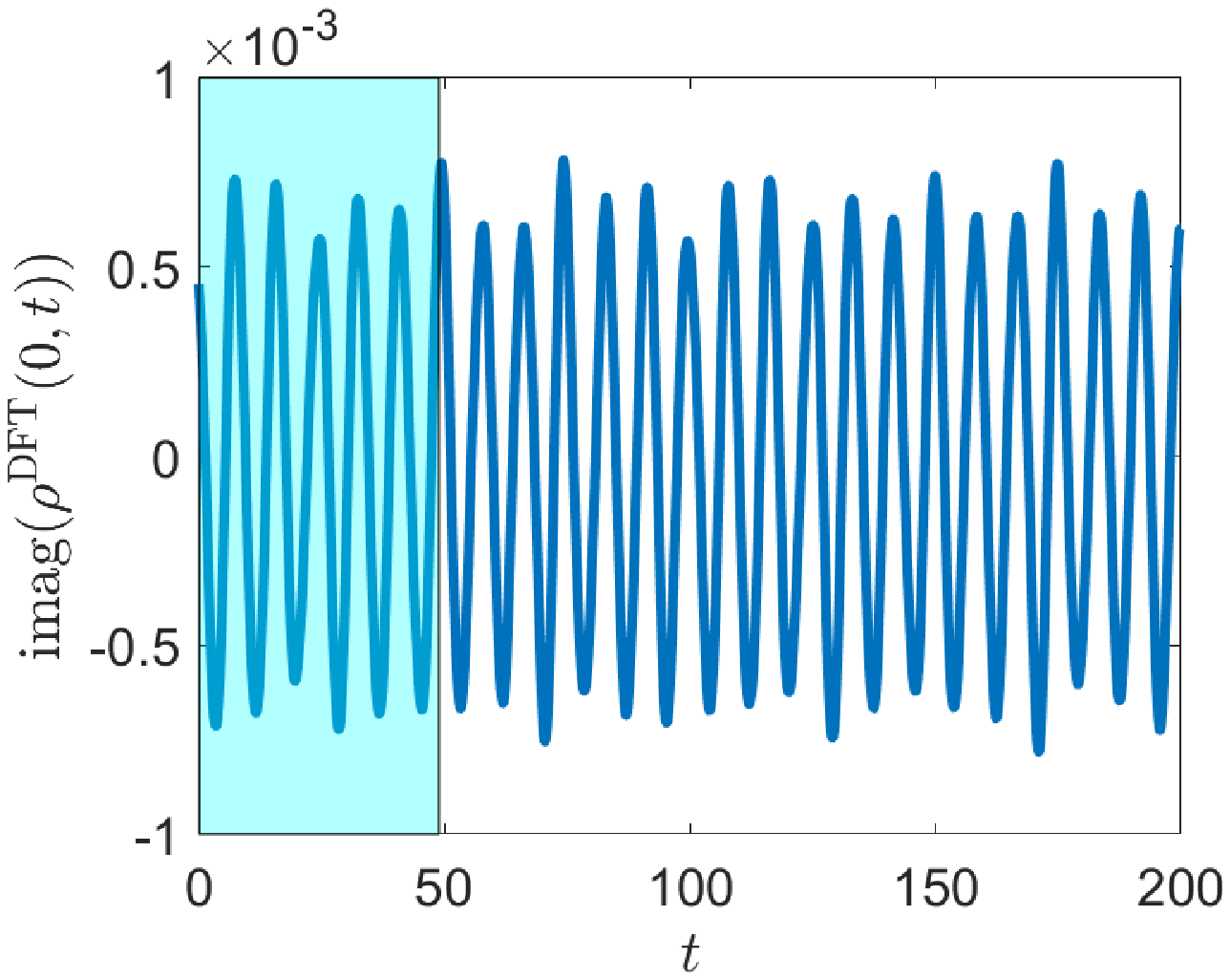}
    \caption{Comparisons of the trajectories at $k = 0$ when light intensity $I = 0.001$. The shaded parts indicate the sampled window of the original trajectory.}
    \label{fig:comp_t_Amp0p001}
\end{figure}

Figure~\ref{fig:comp_t_Amp0p001} shows that, when $I=0.001$, $\rho^{\mathrm{DMD}}$ matches with $\rho$ much better than $\rho^{\mathrm{DFT}}$ at $k=0$. Similar results can be obtained for other $k$-points. In both extrapolations, the number of sampled snapshots is $m=500$, as shown by the shaded window.

To evaluate the quality of reconstruction/extrapolation quantitatively, we introduce a metric defined in terms of the cosine of the angles between the original KBE trajectory vector and the extrapolated trajectory vector at each $k$-point, i.e.,
\begin{equation}
   c_{k}^{\mathrm{a}} = \frac{\langle \rho(k, \cdot),\rho^{\mathrm{a}}(k, \cdot)\rangle}{\|\rho(k, \cdot) \|_{l^2} \|\rho^{\mathrm{a}}(k, \cdot) \|_{l^2}}, \quad k=k_1, ..., k_n,
\label{eq:ckerr}
\end{equation}
where $\mathrm{a}$ is either DMD or DFT, and $\langle \cdot, \cdot \rangle$ denotes the 
standard Euclidean inner product of two complex vectors. It is clear that $|c_k^{\mathrm{a}}|\in[0, 1]$. If the reconstruction/extrapolation $\rho^{\mathrm{a}}(k, \cdot)$ fits the original trajectory $\rho(k, \cdot)$ well, then $|c_k^{\mathrm{a}}|$ should be close to $1$ for all $k=k_1, ..., k_n$. On the contrary, a small value of $|c_k^{\mathrm{a}}|$ suggests a large deviation of $\rho^{\mathrm{a}}(k, \cdot)$ from $\rho(k, \cdot)$ at the $k$-point $k$.

\begin{figure}[htbp]
    	\centering
	\includegraphics[width=0.45\textwidth]{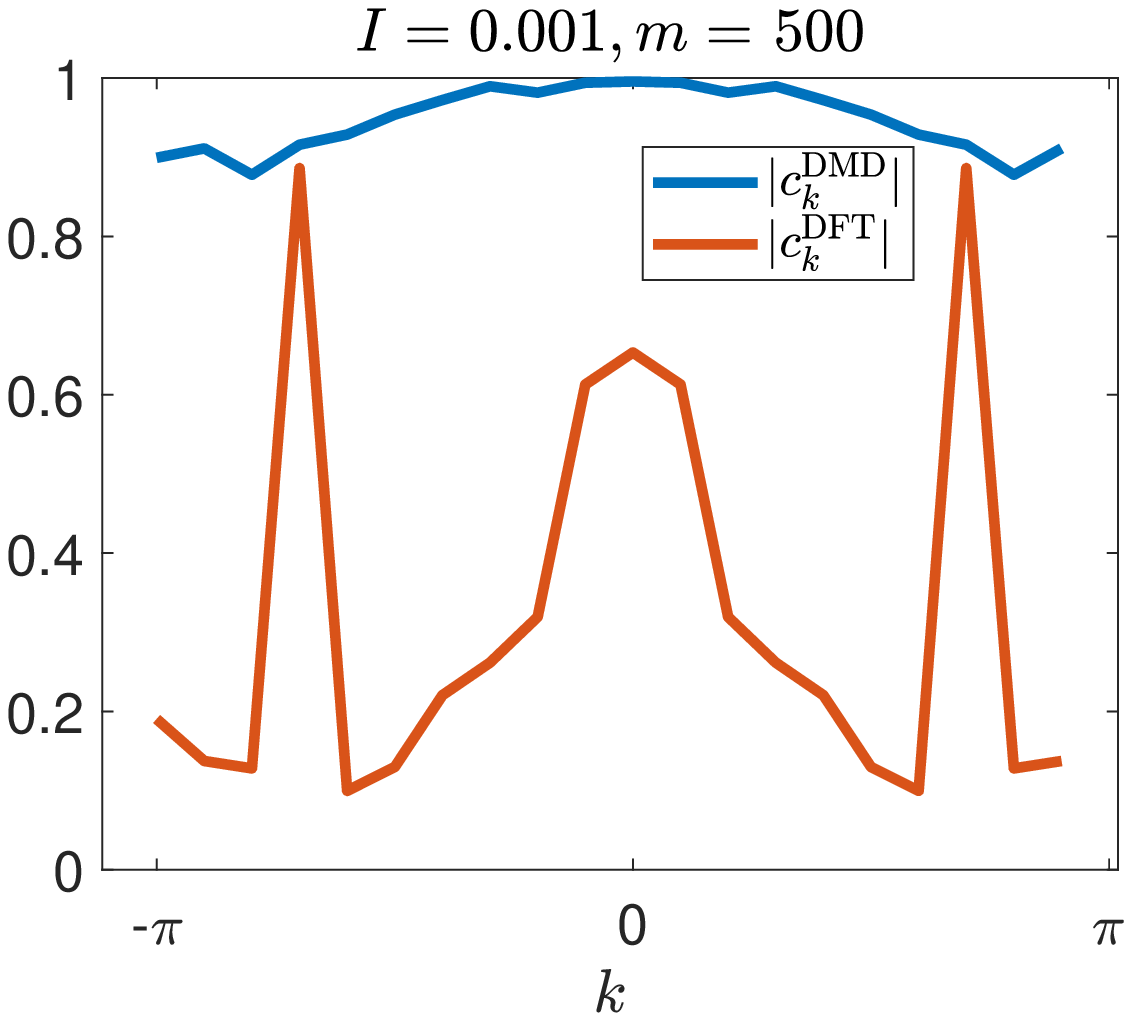}
        \caption{The cosine of the angle between the computed $\rho(k, t)$ from KBE and extrapolated $\rho^{\rm{DMD}}(k, t)$, $\rho^{\rm{DFT}}(k, t)$ when $I = 0.001$.}
        \label{fig:ck0001}
    \end{figure}

Figure~\ref{fig:ck0001} shows that, when $I=0.001$, $|c_{k}^{\mathrm{DMD}}|$ is between $0.82$ and $0.92$ for all $k$-points, indicating a good agreement between the extrapolated trajectory from DMD and the original trajectory. By contrast, $|c_k^{\mathrm{DFT}}|$ is small for most $k$-points, which means the extrapolation from DFT does not give good results. These are consistent with the results in Figure~\ref{fig:comp_t_Amp0p001}.  

\begin{figure}[htbp]
	\centering
	\begin{subfigure}{0.45\textwidth}
		\includegraphics[width=1\textwidth]{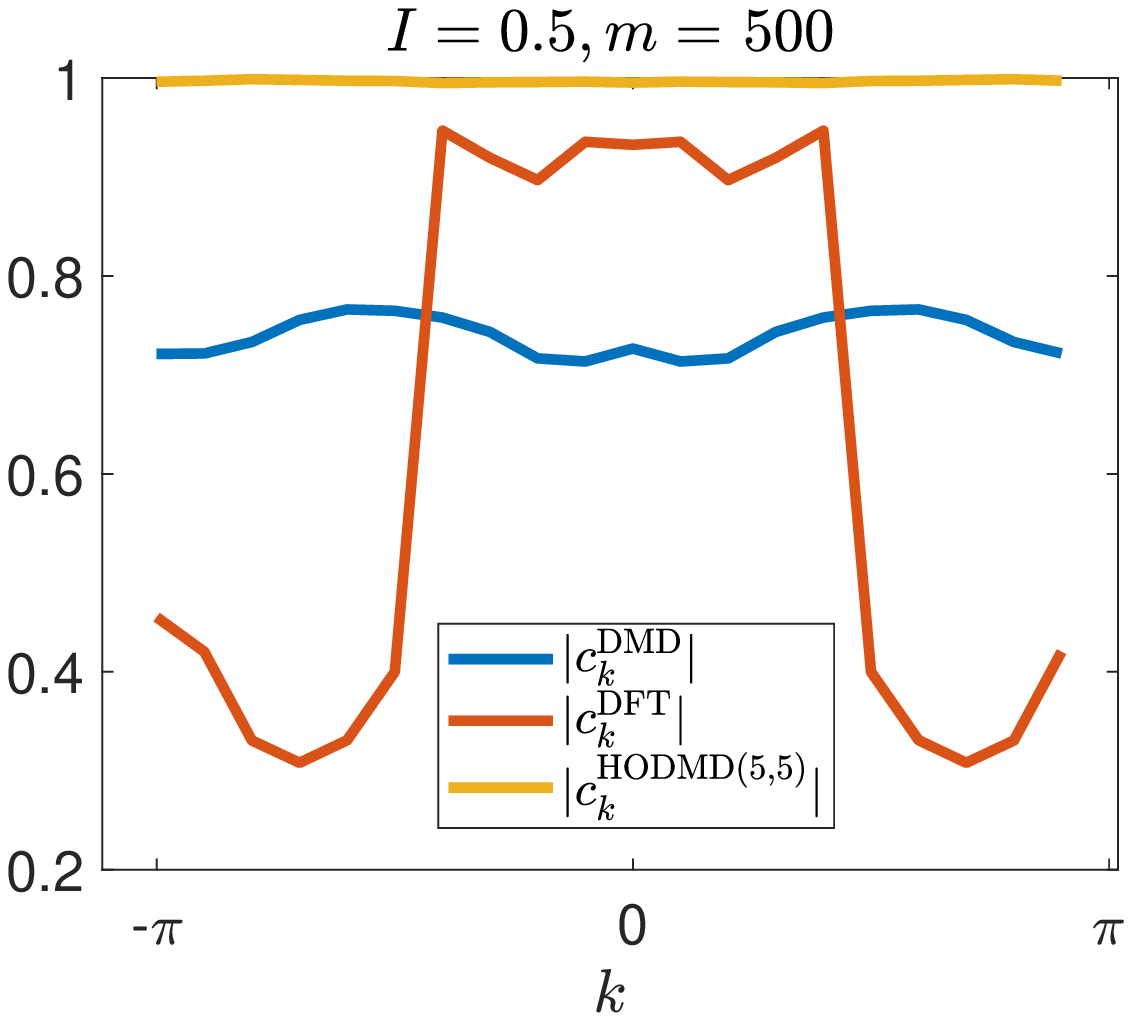}
	\end{subfigure}
	\begin{subfigure}{0.45\textwidth}
		\includegraphics[width=1\textwidth]{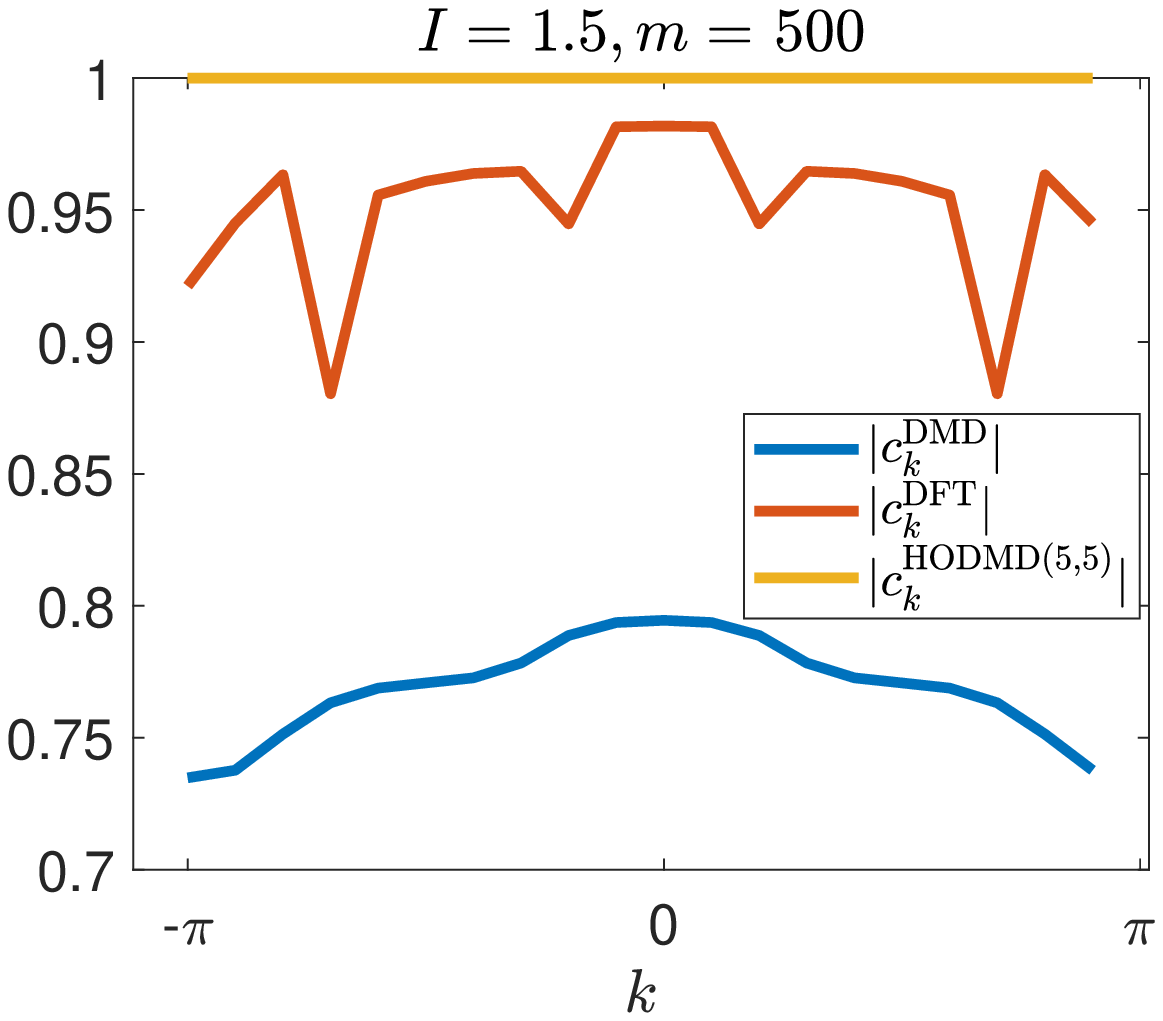}
	\end{subfigure}
    \caption{The cosine of the angle between the computed $\rho(k, \cdot)$ from KBE and extrapolated $\rho^{\rm{DMD}}(k, \cdot)$, $\rho^{\rm{DFT}}(k, \cdot)$, $\rho^{\rm{DMD(5,5)}}(k, \cdot)$ when $I = 0.5$ and $I=1.5$ respectively.}
    \label{fig:ck0p51p5}
\end{figure}

\begin{figure}[htbp]
	\centering
	\includegraphics[width=0.45\textwidth]{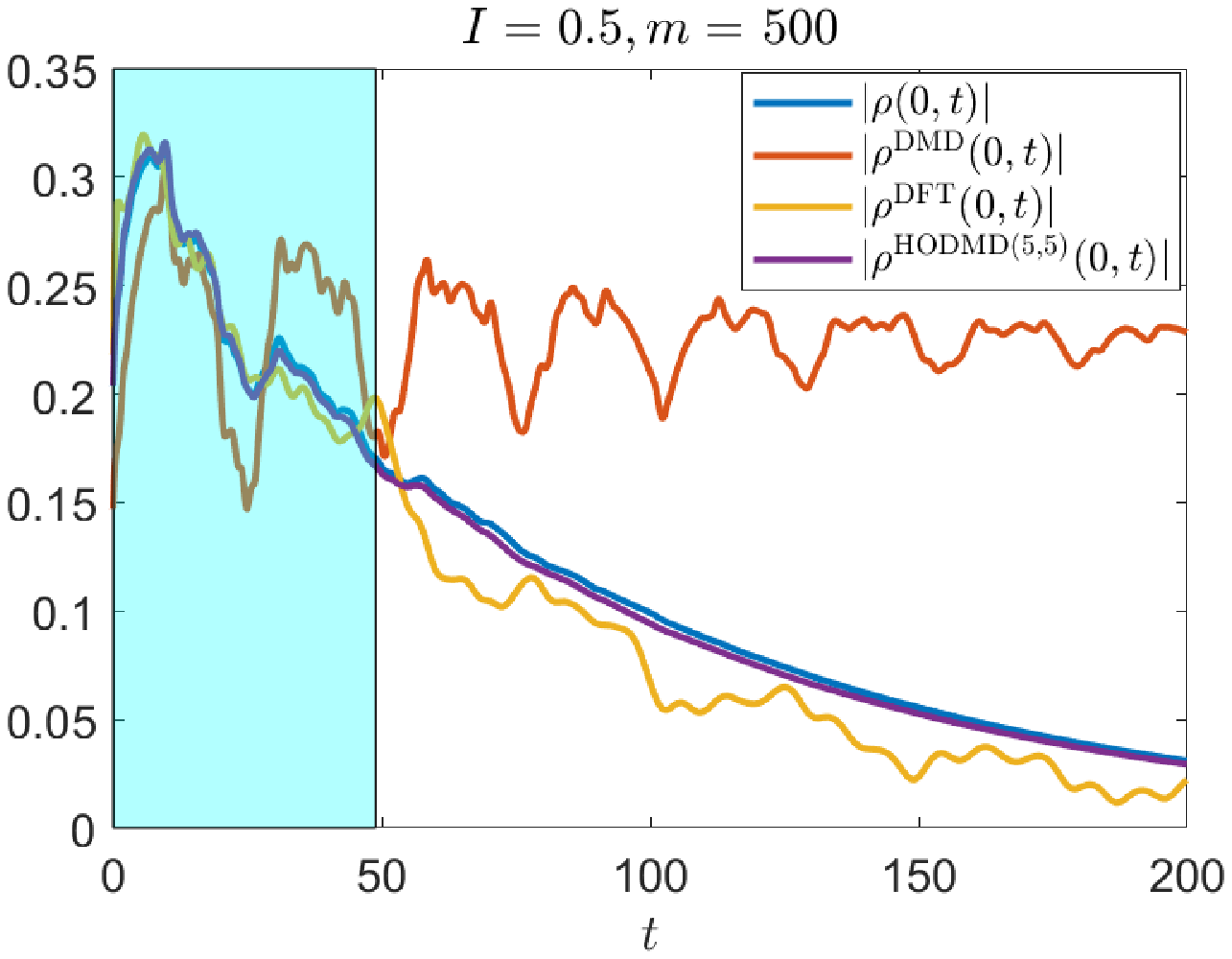}
	\includegraphics[width=0.45\textwidth]{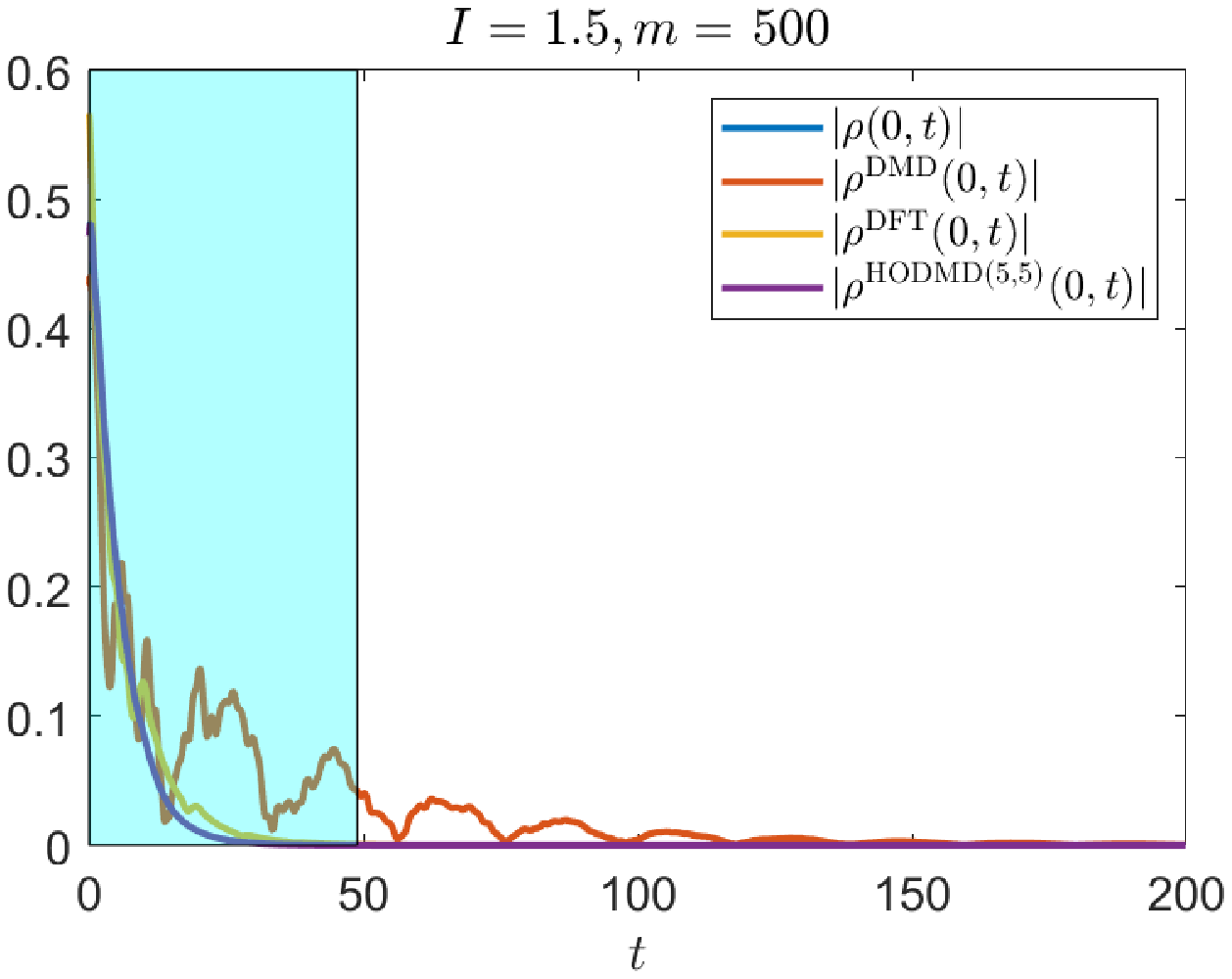}
    \caption{A comparison of $|\rho^{\rm{DMD}}(0, t)|$ , $|\rho^{\rm{DFT}}(0, t)|$ and $|\rho^{\rm{DMD(5,5)}}(0, t)|$ with $|\rho(0, t)|$ for $I=0.5$ and $I=1.5$. The shaded area indicates the sampled window of the trajectories.}
    \label{fig:camp_I0p5_I1p5}
\end{figure}

When the pulse intensity $I$ is increased to $0.5$ or $1.5$, it is observed from Figure~\ref{fig:ck0p51p5} that the values of $|c_k^{\rm{DMD}}|$ is significantly lower. In both cases, $|c_k^{\rm{DMD}}|$ is less than $0.8$ for all $k$-points, which suggests $\rho^{\mathrm{DMD}}(k, \cdot)$ fails to capture the features of $\rho(k, \cdot)$ from the sampled trajectories. Such failure can also be clearly seen in Figure~\ref{fig:camp_I0p5_I1p5} where we plot the magnitude of $\rho^{\mathrm{DMD}}$ and $\rho^{\mathrm{DFT}}$ at the zero $k$-point, and compare them with that of $\rho(t)$.   We remark that when $I=1.5$, although the values of $|c_k^{\rm{DFT}}|$ are large for all $k$-points, there is still noticeable difference between $\rho^{\mathrm{DFT}}(k, \cdot)$ and $\rho(k, \cdot)$ as we can see in Figure~\ref{fig:camp_I0p5_I1p5} within the sampling window.  The reason that the value of $|c_k^{\rm{DFT}}|$ is close to 1.0 in this case is that the inner product between $\rho^{\mathrm{DFT}}(k,t)$ and $\rho(k, t)$ is largely determined by the tails of these two trajectories, which are both nearly zero. In fact, having a $|c_k^{\rm{a}}|$ value close to 1 is a necessary but not sufficient condition for a good extrapolation.
Nonetheless, in this case, the Fourier based extrapolation appears to do a better job in capturing the general trend of the dynamics than the DMD based approach.

We believe the reason why DMD fails to capture the decay rate when $I=0.5$ and $I=1.5$ is that the stronger driving fields introduce more spectral degrees of freedom that are not fully captured by the projected Koopman operator when it is constructed by simply mapping the vector $\rho(t)$  to $\rho(t+\Delta t)$. In other words, as explained in section \ref{sec:hodmd}, there is a discrepancy between the dimension of the projected Koopman operator, i.e., the value of $r$ in \eqref{eq:evol_DMD}, and the intrinsic number of spectral components in the dynamics of $\rho(t)$. 

As we indicated in section \ref{sec:hodmd}, there are two possible ways to address this problem. One is to increase the number of $k$-points. But it is not clear how many $k$-points are needed to ensure the rank of the projected Koopman operator is sufficiently large. In fact, we have tried to increase the number of $k$-points to $n=100$.  That does not appear to yield significant improvement in extrapolation accuracy, but incurs a significant increase in computational and memory cost. This is because the two-time Green's function we need to compute by solving the KBE in order to generate the one time snapshots have more degrees of freedom. 

The other remedy is to use the HODMD algorithm to increase the dimension of the projected Koopman operator by augmenting a single snapshot with $d$ consecutive snapshots as explained in section \ref{sec:hodmd}. These snapshots are concatenated into a single vector and the DMD procedure is then applied to the augmented snapshots \eqref{eq:aug_x}. 

Although it may be possible to estimate the minimum number of time delays to be concatenated into a single column of a snapshot matrix analytically~\cite{Pan2020}, we experimented with several values of $d$ numerically, and found that for the model problem we tested with $I=0.5$ and $I=1.5$, $d = 5$ is a good choice.

Furthermore, as we discussed in section~\ref{sec:hodmd}, to reduce the computational cost of HODMD and the potential level of linear dependency among the column vectors in the snapshot matrix, we can increase the temporal distance between adjacent columns to yield matrices  of the form \eqref{eq:DMDc-d}. Both Figure~\ref{fig:ck0p51p5} and Figure~\ref{fig:camp_I0p5_I1p5} show  that by introducing time delays by 5 $\Delta t$'s and increasing the temporal distance between adjacent columns to 5 $\Delta t$, which yields the HODMD(5,5) scheme, we can obtain extrapolated trajectories that are nearly indistinguishable from the true trajectories (obtained by solving the KBE numerically) even when the driving field intensities are increased to $I=0.5$ and $I=1.5$ respectively. 
  
\begin{figure}[htbp]
	\centering
	\begin{subfigure}{0.45\textwidth}
		\includegraphics[width=1\textwidth]{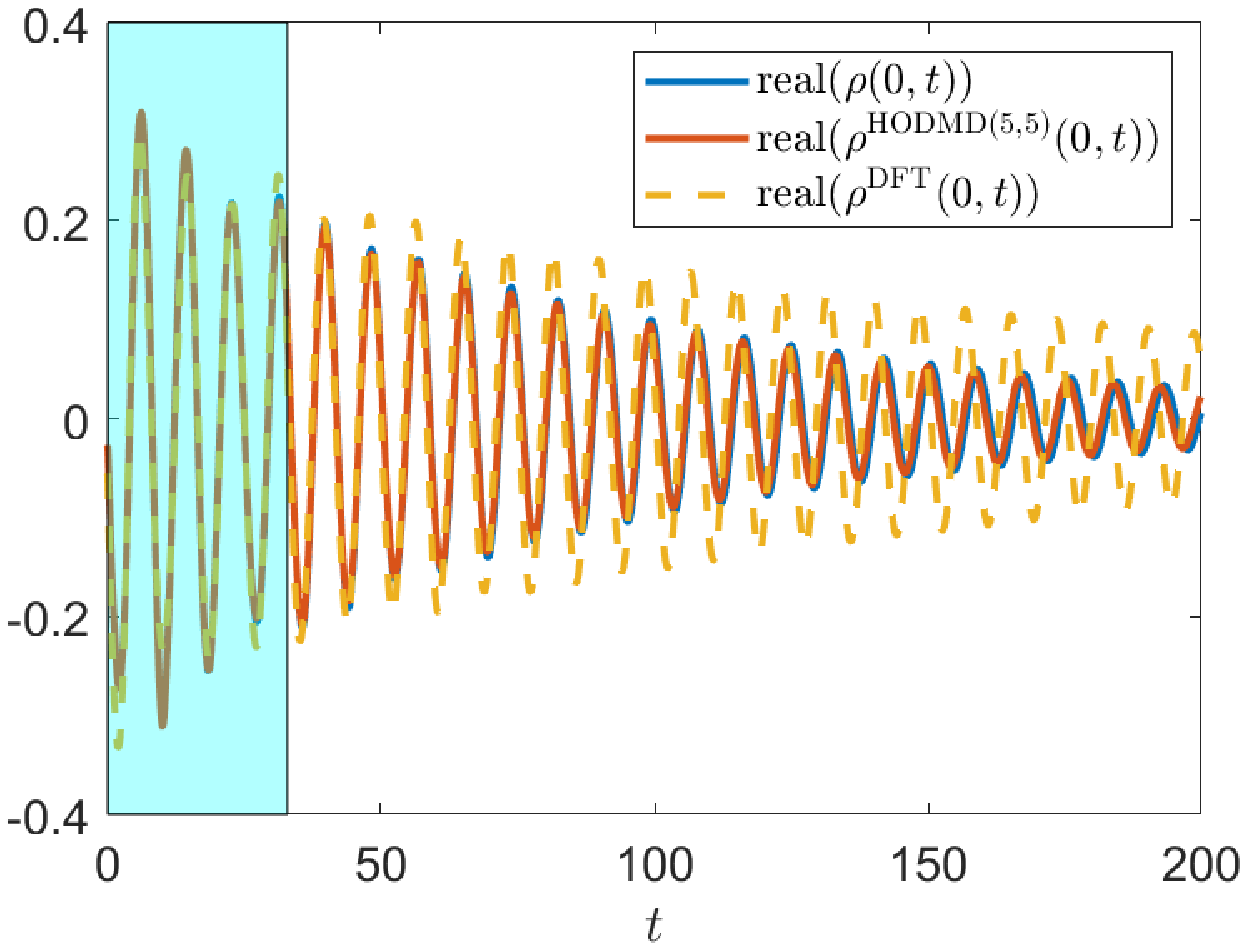}
	\end{subfigure}
	\begin{subfigure}{0.45\textwidth}
		\includegraphics[width=1\textwidth]{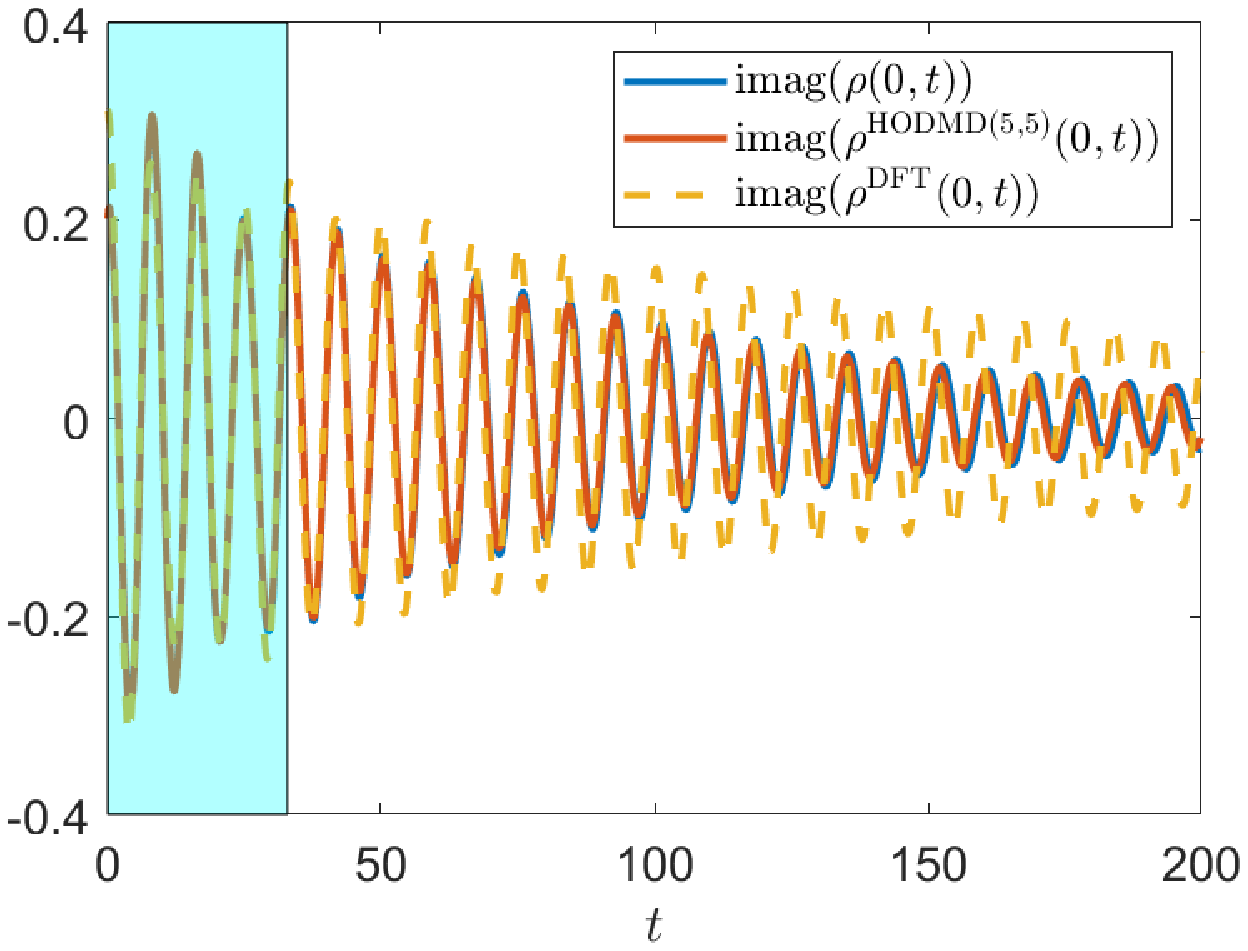}
	\end{subfigure}
	\caption{Comparisons of the trajectories at $k=0$ when the driving intensity $I = 0.5$. The trajectories are reconstructed and extrapolated with $m=340$. The shaded parts demonstrate the sampled window of the original trajectory.}
	\label{fig:comp_t_Amp0p5}
\end{figure}

\begin{figure}[htbp]
	\centering
	\begin{subfigure}{0.45\textwidth}
		\includegraphics[width=1\textwidth]{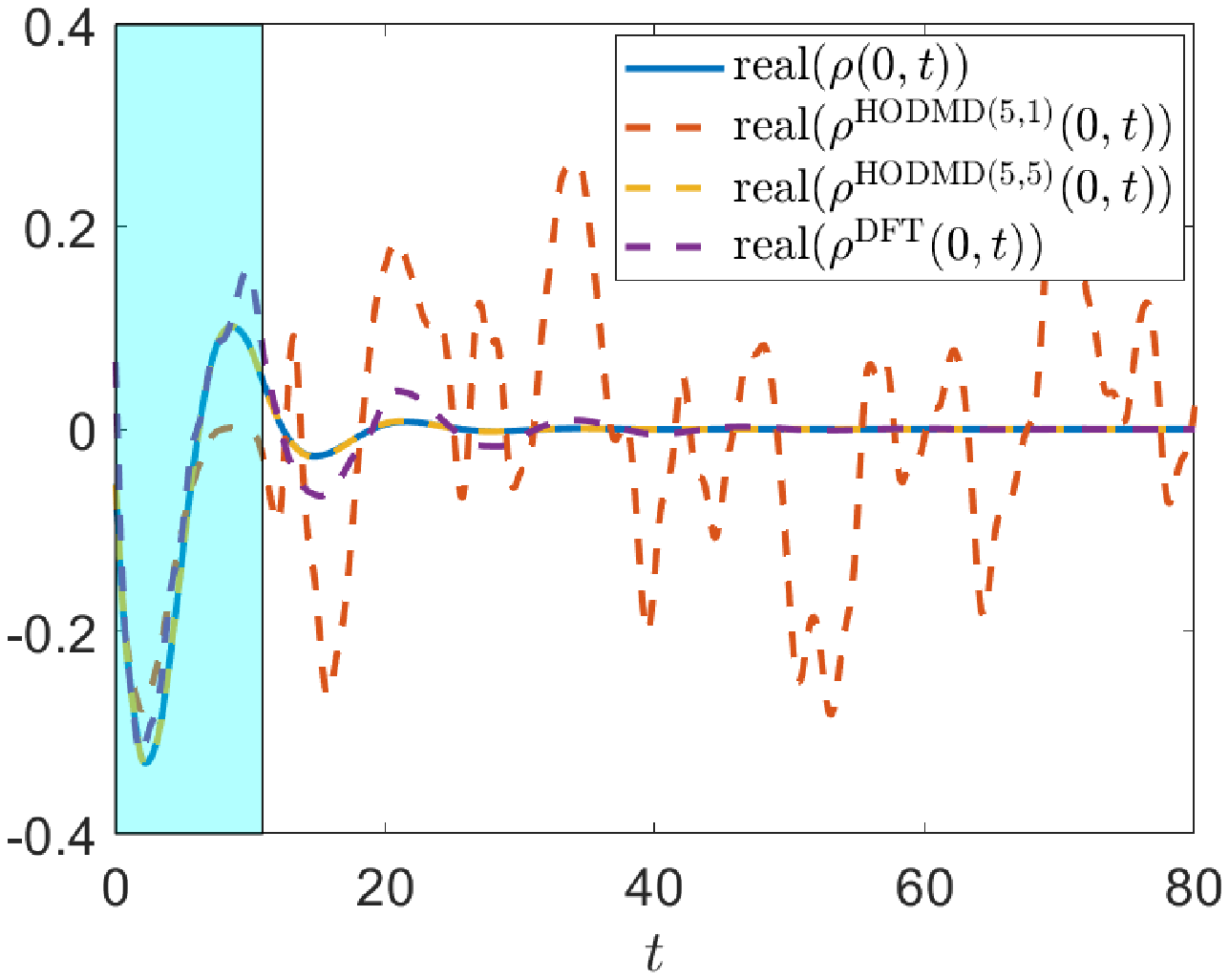}
	\end{subfigure}
	\begin{subfigure}{0.45\textwidth}
		\includegraphics[width=1\textwidth]{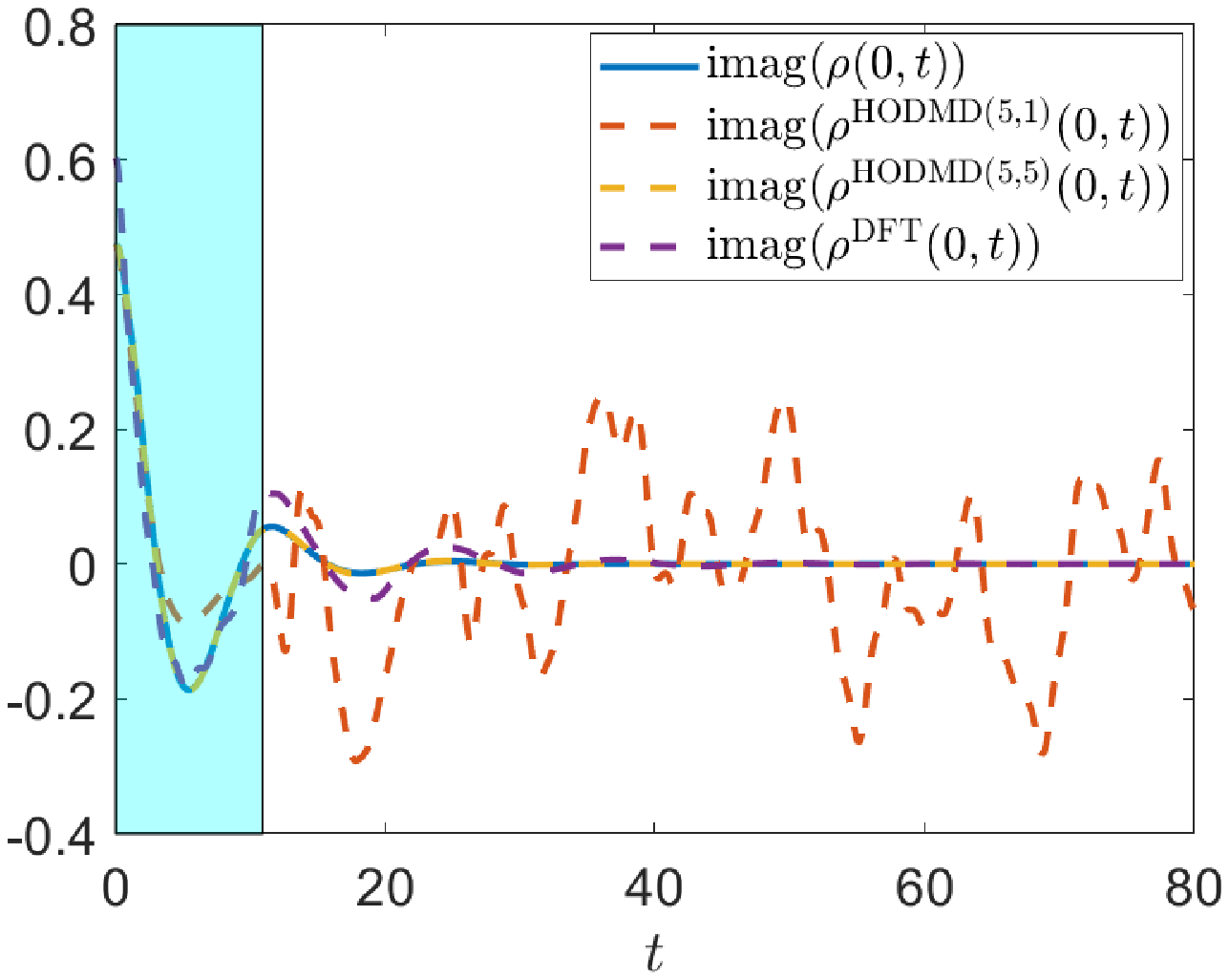}
	\end{subfigure}
	\caption{Comparisons of the trajectories at $k=0$ when the driving intensity $I = 1.5$. The trajectories are reconstructed and extrapolated with $m=120$. The shaded parts demonstrate the sampled window of the original trajectory.}
	\label{fig:comp_t_Amp1p5}
\end{figure}

Figures~\ref{fig:comp_t_Amp0p5} and~\ref{fig:comp_t_Amp1p5} 
show that with a reduced number of sampled snapshots, the extrapolated trajectories produced by HODMD(5,5) still match perfectly with the true trajectories associated with the driving field intensity $I=0.5$ and $I=1.5$ respectively.  They are clearly better than the trajectories produced from the DFT based extrapolation, which fails to accurately capture the decay rate of the dynamics. Moreover, the computational cost of the DFT based extrapolation is much higher because a nonlinear least squares optimization problem needs to be solved in order to determine the coefficients in the extrapolation model. The solution of the optimization problem depends sensitively on the initial guess of the coefficients.

In addition, it can be clearly seen from Figure~\ref{fig:comp_t_Amp1p5} that when $I=1.5$, HODMD($5,1$) fails to extrapolate the trajectory correctly with $m=120$ snapshots. The failure is likely due to the linear dependency among columns of the $\mathbf{\tilde{X}}_1$ snapshot matrix, and a less optimal singular value threshold used in the truncated SVD performed on this matrix. 

To demonstrate that HODMD indeed captures the fast decay of $\rho(t)$ than DMD, we list the DMD and HODMD(5,5) frequencies associated with the four most significant modes in  Table~\ref{tab:freq_Amp0p5} for the simulation with $I=0.5$, and Table~\ref{tab:freq_Amp1p5} for the simulation with $I=1.5$.  \begin{table}[htbp]
    \centering
    \begin{tabular}{|c|c|c|c|c|}
    \hline
       & 1 & 2 & 3 & 4\\
    \hline
      $\omega^{\rm{DMD}}$ & $0.727$ & $2.804+0.015i$ & $2.190+0.012i$ & $3.419+0.019i$ \\
    \hline
      $\omega^{\rm{HODMD(5,5)}}$ & $0.741+0.012i$ & $2.208+0.059i$ & $2.826+0.064i$ & $3.455+0.073i$ \\
    \hline
    \end{tabular}
    \caption{The four most significant frequencies from DMD and HODMD(5,5) when the driving field intensity $I=0.5$ with $m=500$.}
    \label{tab:freq_Amp0p5}
\end{table}

\begin{table}[htbp]
    \centering
    \begin{tabular}{|c|c|c|c|c|}
    \hline
       & 1 & 2 & 3 & 4\\
    \hline
      $\omega^{\rm{DMD}}$ & $-0.581+0.226i$ & $0.409+0.030i$ & $0.712+0.029i$ & $2.935+0.049i$ \\
    \hline
      $\omega^{\rm{HODMD(5,5)}}$ & $0.487+0.193i$ & $0.746+0.245i$ & $2.400+0.293i$ & $3.046+0.323i$ \\
    \hline
    \end{tabular}
    \caption{The four most significant frequencies from DMD and HODMD(5,5) when the driving field intensity $I=1.5$ with $m=500$.}
    \label{tab:freq_Amp1p5}
\end{table}

From these tables, we observe that, overall, the imaginary parts of $\omega^{\rm{HODMD(5,5)}}$ are much larger than those of $\omega^{\rm{DMD}}$. By the convention we use in \eqref{eq:dmd_recon}, a large imaginary part corresponds to a more rapid decay of the DMD or HODMD mode. 

One practical question one may ask is how many snapshots ($m$) we need to collect in order to accurately extrapolate the full (long-time) trajectory of $\rho(t)$. This question is difficult to answer a priori in general. For the model problem we examined, we experimented numerically with $m$ values ranging from 21 to 500.  For each $m$, we  
compute the root mean square error of the extrapolated trajectory $\rm{err}_{\rm{b}}$ defined as
\begin{equation}
	\text{err}_{\text{b}} = \left[\frac{\sum_{j=1}^n\|\rho(k_j, t_{m+1}:t_{N}) - \rho^{\text{b}}(k_j, t_{m+1}:t_{N})\|_{l^2}^2}{N-m}\right]^{1/2}, \quad \text{b} = \text{HODMD($d,s$)}.
\end{equation}

\begin{figure}[htbp]
	\centering
	\begin{subfigure}{0.45\textwidth}
		\includegraphics[width=1\textwidth]{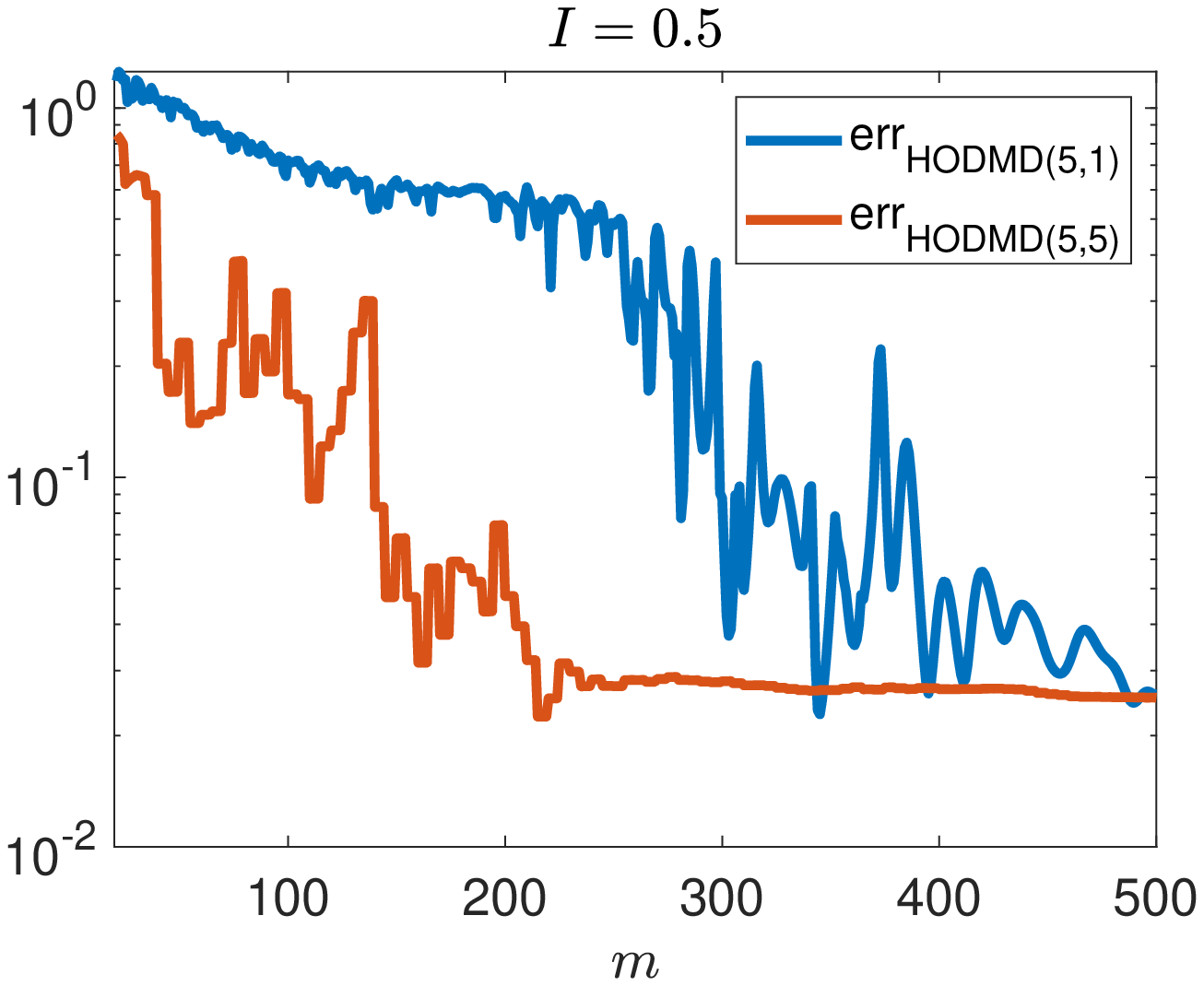}
	\end{subfigure}
	\begin{subfigure}{0.45\textwidth}
		\includegraphics[width=1\textwidth]{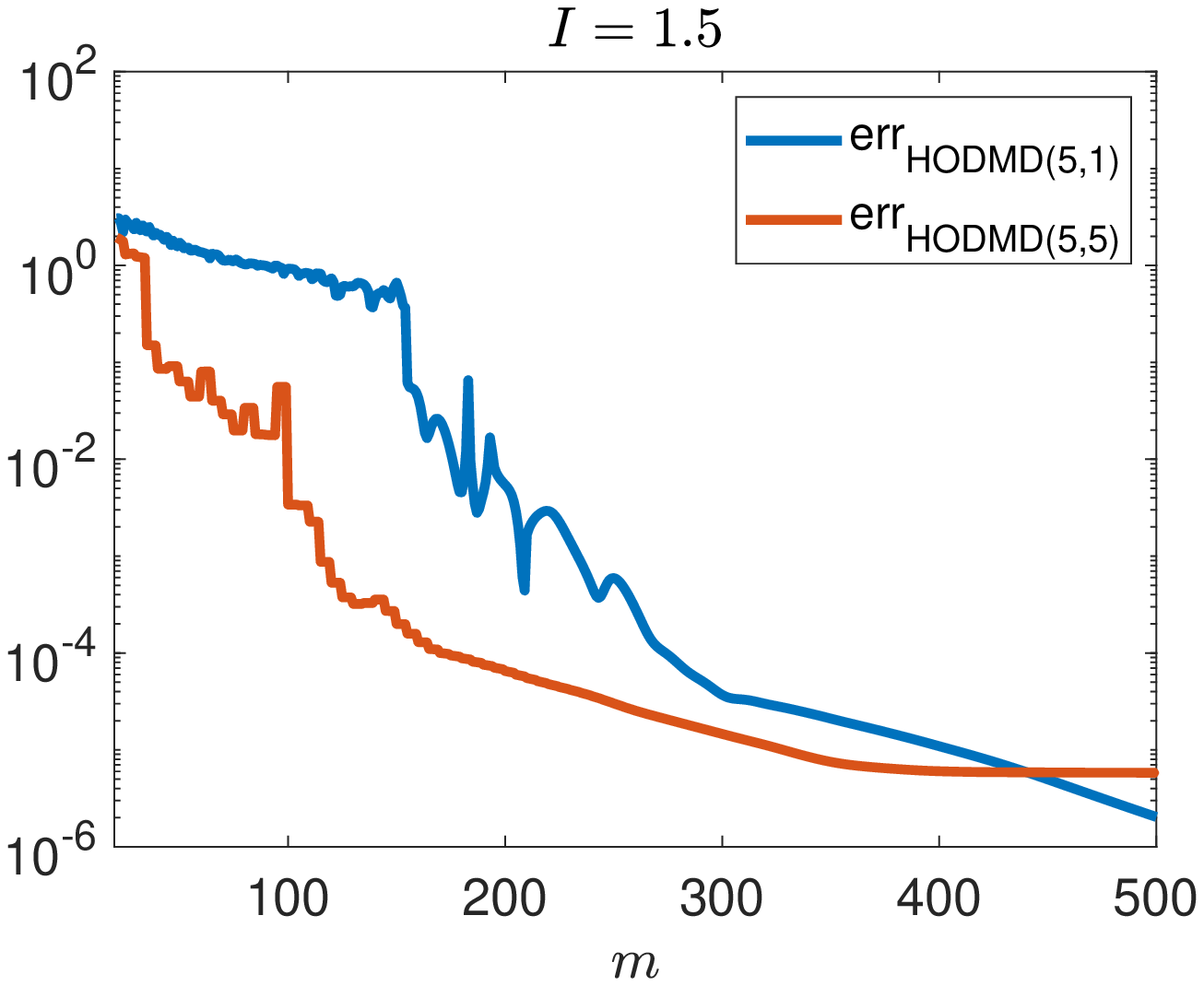}
	\end{subfigure}
	\caption{Comparisons of the root mean square errors of the extrapolated trajectories from HODMD(5,1) and HODMD(5,5), when the driving intensity $I = 0.5$ and $I=1.5$.}
	\label{fig:comp_err}
\end{figure}

In Figure~\ref{fig:comp_err}, we plot this error associated with both HODMD$(5,1)$ and HODMD$(5,5)$ for these $m$ values. 
For both test problems (with $I=0.5$ and $I=1.5$), we observe that the extrapolation error generally decreases as $m$ increases, which is expected. The error clearly decreases faster for HODMD(5,5) than for HODMD(5,1).
Because the time step size used in the Runge-Kutta scheme to solve the KBE is $\Delta t=0.1$. The true trajectory is only accurate up to $O(10^{-2})$. Therefore, approximately 200 snapshots are sufficient to produce an accurate HODMD(5,5) extrapolation when $I=0.5$. Far fewer snapshots are needed for $I=1.5$. This is mainly due to the fact the dynamics has more features in early time and quickly decays to zero as $t$ increases.  

\section{Conclusion}
\label{sec:conclude}

We applied DMD to study the dynamics of a one-time physical observable originated from the simulation of a many-body system away from equilibrium through the Green's function approach. Traditional numerical methods to obtain the observable involve solving coupled two-time nonlinear integral differential equations, which results in high memory requirement and large computational cost. In contrast, the data-driven DMD method only depends on a small sampled set of the numerical solutions, and can be easily applied through truncated SVD decomposition.

Numerical results obtained from the dynamical simulation of a two-band model system show that DMD successfully captures the major dynamical modes and frequencies of the observable when there is no or little external light-matter interaction. In these equilibrium or near equilibrium cases, the DMD modes are physical as can be verified by performing a linear response analysis and comparing the DMD modes with the eigenvalues and eigenvectors of the corresponding BSE Hamiltonian.
When the driving pulse intensity in the interaction term of the two-band model is large, the standard DMD fails to accurately reconstruct or extrapolate the nonequilibrium dynamics, because the rank of the projected Koopman operator is too small. Under this circumstance, we introduced HODMD($d,s$) with concatenated snapshots, which is derived from the augmented Koopman operator. Numerical examples show that HODMD($d,s$) perfectly solves the problem. Moreover, increasing $s$ in HODMD($d,s$) can improve the efficiency and accuracy of the algorithm as the data matrix becomes more compact, and there is less linear dependency among the augmented snapshots. Compared to Fourier analysis, DMD and HODMD($d,s$) can perform the reconstruction and extrapolation of the original trajectories more accurately with less computational cost.

\section*{Acknowledgments}
This work is supported by the Center for Computational Study of Excited-State Phenomena in Energy Materials (C2SEPEM) at the Lawrence Berkeley National Laboratory, which is funded by the U.\,S. Department of Energy, Office of Science, Basic Energy Sciences, Materials Sciences and Engineering Division, under Contract No. DE-AC02-05CH11231, as part of the Computational Materials Sciences Program. 
The authors acknowledge the computational resources of the National Energy Research Scientific Computing (NERSC) center.

\bibliographystyle{abbrv}
\bibliography{ref} 

\end{document}